\documentclass[fleqn,usenatbib]{mnras}

\usepackage{newtxtext,newtxmath}

\usepackage[T1]{fontenc}
\usepackage{ae,aecompl}

\usepackage{graphicx}	
\usepackage{amsmath}	
\usepackage{multirow}
\usepackage{enumitem}
\usepackage{ulem}

\title[Dust Condensation]{Dust Condensation in Evolving Discs and the Composition of Planetary Building Blocks}

\author[M. Li et al.]{
Min Li,$^{1}$
Shichun Huang,$^{2}$
Michail I. Petaev,$^{3,4}$
Zhaohuan Zhu,$^{1}$
and Jason H. Steffen$^{1}$\thanks{E-mail: jason.steffen@unlv.edu}
\\
$^{1}$Department of Physics and Astronomy, University of Nevada, Las Vegas, 4505 S. Maryland Pkwy, Las Vegas 89154, USA\\
$^{2}$Department of Geoscience, University of Nevada, Las Vegas, 4505 S. Maryland Pkwy, Las Vegas 89154, USA\\
$^{3}$Department of Earth \& Planetary Sciences,Harvard University, 20 Oxford St., Cambridge 02138, USA\\
$^{4}$Harvard-Smithsonian Center for Astrophysics, 60 Garden St., Cambridge 02138, USA
}

\date{Accepted XXX. Received YYY; in original form ZZZ}

\pubyear{2020}

\begin{document}
\label{firstpage}
\pagerange{\pageref{firstpage}--\pageref{lastpage}}
\maketitle

\begin{abstract}
Partial condensation of dust from the Solar nebula is likely responsible for the diverse chemical compositions of chondrites and rocky planets/planetesimals in the inner Solar system.  We present a forward physical-chemical model of a protoplanetary disc to predict the chemical compositions of planetary building blocks that may form from such a disc.  Our model includes the physical evolution of the disc and the condensation, partial advection, and decoupling of the dust within it.  The chemical composition of the condensate changes with time and radius.  We compare the results of two dust condensation models: one where an element condenses when the midplane temperature in the disc is lower than the 50\% condensation temperature ($\rm T_{50}$) of that element and the other where the condensation of the dust is calculated by a Gibbs free energy minimization technique assuming chemical equilibrium at local disc temperature and pressure.  The results of two models are generally consistent with some systematic differences of $\sim 10$\% depending upon the radial distance and an element's condensation temperature.  Both models predict compositions similar to CM, CO, and CV chondrites provided that the decoupling timescale of the dust is on the order of the evolution timescale of the disc or longer.  If the decoupling timescale is too short, the composition deviates significantly from the measured values.  These models may contribute to our understanding of the chemical compositions of chondrites, and ultimately the terrestrial planets in the solar system, and may constrain the potential chemical compositions of rocky exoplanets.
\end{abstract}




\section{Introduction}

\begin{figure*}
	\includegraphics[width=1.9\columnwidth]{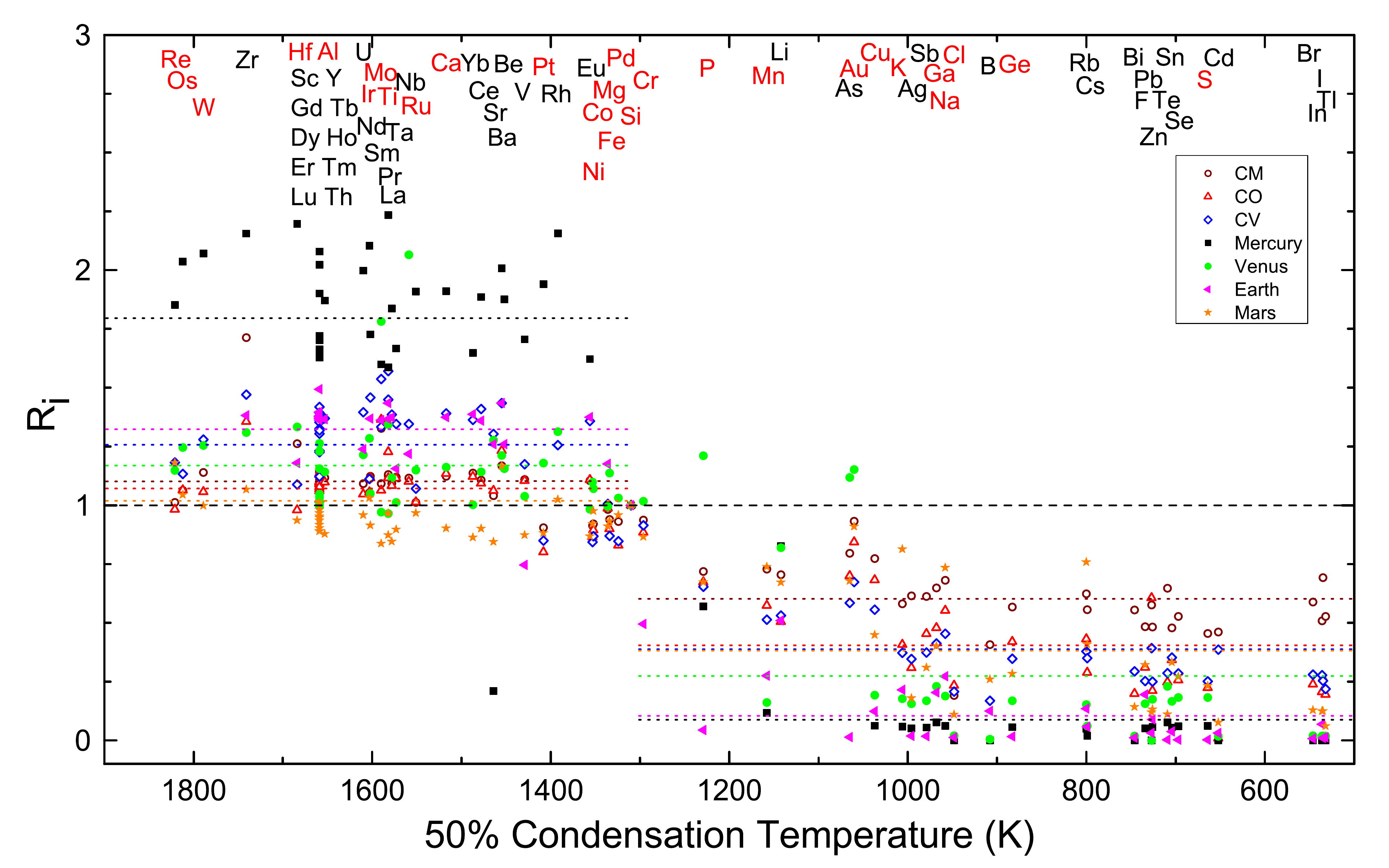}
    \caption{\textbf{Observed elemental abundances: } Relative abundances, $\rm R_i$, in terrestrial planets and CM, CO, and CV chondrites,  normalized to the CI chondrites and Si, as a function of $\rm T_{50}$ \citep{Lodders:2003}.  Data sources: chondrites \citep{Wasson:1988}, bulk silicate Mercury \citep{Morgan:1980}, bulk silicate Venus \citep{Morgan:1980}, bulk silicate Earth \citep{McDonough:1995}, and bulk silicate Mars \citep{Lodders:1997}.  The horizontal line at 1 indicates where a body is neither enriched nor depleted in a given element relative to Si.  Horizontal dashed lines show the median depletion in these objects at $\rm T_{50}$ above and below 1310K. This figure shows that all terrestrial planets and chondrites are depleted in volatiles, and are consequently enriched in refractory elements relative to the Sun.  Highly siderophile elements in Earth are not plotted because they were affected by core formation. Elements in red font are modeled with the GRAINS code (See Section \ref{sec:grain}).  
   }
    \label{fig:all-ele}
\end{figure*}

Our Solar system is an important testbed for studying the process of planet formation.  The chemical composition of terrestrial planets and both differentiated and undifferentiated meteorites can constrain its formation and evolution.  All terrestrial planets and many carbonaceous chondrites (CM, CO, and CV) are depleted in volatile elements and enriched in refractory elements relative to the composition of the Sun, as represented by CI chondrites \citep{Asplund:2005,Asplund:2009}.  The degree of depletion of these elements is correlated with their volatility \citep{Palme:2014}.

Figure \ref{fig:all-ele} shows the relative elemental abundances of terrestrial planets and a number of carbonaceous chondrites normalized to the Solar abundance and Si
\begin{equation}\label{equ.ri}
{\rm R_i=\frac{[\chi_{E}/\chi_{Si}]_P}{[\chi_{E}/\chi_{Si}]_{\odot}}},
\end{equation}
as a function of the ``50\% condensation temperatures'', $\rm T_{50}$, for each element as tabulated in \citet{Lodders:2003}.  The 50\% condensation temperature is the temperature where half of a given element is in the condensed phase under a total pressure of $10^{-4}$ bar with Solar composition.  Here, $\chi_{\rm E}$ and $\chi_{\rm Si}$ stand for the concentrations of an element (E) and Si, respectively, and subscript P refers to a planet or chondrite.  $\rm R_i$ for terrestrial planets and CM, CO, and CV chondrites are smaller than unity for volatile elements while larger than unity for refractory elements.  Determining causes of volatile element depletion in rocky planets, as well as the role that the physical and chemical environment plays in producing this depletion, is crucial for our understanding of the origin of our Solar system and is the subject of ongoing research in planetary science.

Since planetesimals and planets form in protoplanetary discs, their composition should be a function of both the evolution of the gaseous disc and the condensation of the elements in it.  
Since the early 1970s, chondrites and chondritic components such as CAIs (Calcium-Aluminium-rich Inclusions) are hypothesized to form from the hot solar nebula during the early stage of the Solar system formation.  To understand how this formation proceeds, one must model both the physical and chemical processes involved. 

\subsection{Previous work}

Partial condensation models are favored to explain the volatile depletion in rocky planets and many carbonaceous chondrites (CM, CO, and CV).  \citet{Grossman:1972} was among the first to use a thermodynamic approach to study the chemical evolution of the Solar nebula and the condensation of dust.  He suggested that the CAIs are direct condensates from the nebula.  Other early calculations were performed by \citet{Urey:1955}, \citet{Lord:1965}, and \citet{Larimer:1967}.  \citet{Lewis:1974} applied the same concept to understand the chemical composition of the terrestrial planets.

An important milestone for our purposes is \citet{Cassen:1996}, who modeled the chemical evolution of an evolving disc to explain the volatile element depletion in CM, CO, and CV chondrites.  He assumed that an element begins to condense once the temperature at a given radius in the disc drops below $\rm T_{50}$ (estimated at a pressure of $10^{-4}$ bar).  Despite this simplifying assumption, he was able to broadly reproduce the volatile element depletions in carbonaceous chondrites by forming them between 1-3 AU.  \citet{Cassen:2001} used a similar model to calculate the relative abundance of the elements in the large rocky materials at different radii.  The main differences in this case were the inclusion of two-dimensional temperature distributions within the disc, external illumination, and the migration of the coagulated solids toward the Sun.

Later, \citet{Ciesla:2008} investigated whether the cooling disc can explain the depletion of moderately volatile elements (with condensation temperature between $\sim$ 650 - 1350 K) in chondritic meteorites.  Compared to \citet{Cassen:1996}, he calculated the viscous evolution of the disc and included the effects of gas drag on migrating pebbles.   
He found that the observed depletion can be explained by his model within a small range of parameters.  In order to reproduce the depletion pattern observed in the asteroid belt, the initial temperature of the disc needed to be higher than roughly 1350 K beyond 2 AU and  the formation timescale of planetesimals should be less than $10^5$ yr.  From this result it remains challenging to explain the observations that chondrule formation began a few million years after CAI formation \citep{Kita:2000,Amelin:2002,Connelly:2012}.

Other groups considered the chemical equilibrium of the system.  Using the commercial software package HSC Chemistry, \citet{Bond:2010a} calculated the equilibrium partitioning of 16 elements among 78 gaseous and 33 solid species at seven specific times in the evolving disc.  They then modeled the formation of the terrestrial planets using N-body accretion simulations \citep{OBrien:2006}.  Using the same method but different initial chemical compositions of host stars, \citet{Bond:2010b} also modeled the formation of extrasolar planets -- showing that the resulting compositions of exoplanets are diverse, depending upon elemental abundances of the host star, ranging from Earth-like to carbide-rich planets.  

Using the HSC Chemistry software, three different disc models, and elemental abundances of the present-day Solar photosphere \citep{Asplund:2005}, \citet{Elser:2012} modeled the composition of the terrestrial planets.  They adopted a self-consistent approach to determine the time in the evolving disc for calculating radial variations in disc chemistry assuming chemical equilibrium.  They showed that 
disc models with high temperatures and pressures produce a large variety in the composition of the resulting planets.  \citet{Pignatale:2016} adopted a two dimensional disc model and used the same software to model 170 gas and 317 solid compounds that form from 15 elements.  They used a snapshot of the disc at 1 Myr as a proxy for typical disc conditions (temperature and pressure in the midplane of the disc) and modeled chemical and mineralogical gradients in the disc.

Given this background, none of these previous simulations consider simultaneous physical and chemical disc evolution or how differential transport of dust and gas affect dust condensation calculation.  As the disc evolves with time, the surface density, temperature, and pressure in the disc change, which affects the condensation of chemical elements.  Therefore, a time-dependent disc model is needed which considers the effects of changing pressure and surface density on the condensation temperatures as a function of time and radial distance of the evolving disc.  

In this work, we present composition results for a dynamically evolving disc that incorporates real-time dust condensation.  We use two different models.  The first is the same method as \citet{Cassen:1996} but with updated T$_{50}$s.  The second is a modified version of the evolving disc from \citet{Cassen:1996} and to use GRAINS code from \citet{Petaev:2009} to model the chemical equilibrium.

Previous work by \citet{Moriarty:2014} did include a model that took into account both the evolution of disc and the disc chemistry, using the same chemical equilibrium calculation as \citet{Bond:2010a,Bond:2010b} and modeling the disc and dust evolution in a fashion similar to our approach.  Their motivation was to seed the initial composition of a planetesimal disc to model terrestrial planet formation in various extrasolar systems with different chemical compositions, rather than to compare with solar system chondrites.  They found a good agreement with previous studies for Sun-like stars with C/O $\sim 0.54$ and predicted a number of carbon-rich planets around stars with C/O $> 0.65$.  We build upon their approach by including a larger list of chemical elements and compounds (33 vs 16 elements and 762 vs 111 species respectively).  In addition, we examine the effects of changing the assumptions on the disc evolution and dust condensation timescales.

\begin{table}
	\caption{The 50\% condensation temperatures from different models}  
	\label{tab:element}
	\setlength{\tabcolsep}{1mm}{
	\begin{tabular}{ccccc} 
		\hline\hline
		 \multirow{2}{*}{element} &
		\multicolumn{3}{c}{50\% Condensation Temperature}& \multirow{2}{*}{GRAINS T$_2^{(\rm d)}$} \\
        \cline{2-4}
        & GRAINS$^{(\rm a)}$ & Lodders$^{(\rm b)}$   &   Cassen$^{(\rm c)}$  \\
        \hline
		H & ...  & ... & ...& ...\\
		He & ... & ... & ...& ... \\
		C & ...  & 40 & ... & ...\\
		N & ...  & 123 & ... & ...\\
		O & ...  & 180 & ...& ... \\
		Na & 988  & 958 & 971 & 1069\\
		Mg & 1336  & 1336 & ...& 1385 \\
		Al & 1654  & 1653 & ...& 1675 \\
		Si & 1316  & 1310 & 1311 & 1478\\
		P & 1318  & 1229 & 1151& 1358 \\
		S & 669  & 664 & 652 & 711\\
		Cl & 440  & 948 & ...& 446 \\
		K &906  & 1006 &	1002 & 1014\\
		Ca & 1500  &	1517 & ... & 1669\\
		Ti & 1583  &	1582 & ...& 1594 \\
		Cr & 1304  &	1296 & 1276& 1358 \\
		Mn & 1154  &	1158 & 1191& 1263 \\
		Fe & 1335  &	1334 &	... & 1360\\
		Co & 1350  &	1352 &	...& 1361 \\
		Ni & 1264  &	1353 &	1354& 1326 \\
		Cu & 1100  &	1037 &	1038 & 1232\\ 
		Ga & 1018  &	968 &	920 & 1159 \\
		Ge & 911  &	883 &	826& 1062 \\
		Mo & 1548  &	1590 &	...& 1644 \\ 
		Ru & 1544  &	1551 &	...& 1630 \\
		Pd & 1336  &	1324 &	1334& 1360 \\
		Hf & 1730  &	1684 &	...& 1749 \\
		W &	1788  &	1789 &	... & 1832\\
		Re & 1770  &	1821 &	...& 1830 \\
		Os & 1809  &	1812 &	...& 1833 \\
		Ir & 1584  &	1603 &	...& 1691 \\
		Pt & 1396  &	1408 &	... & 1509\\ 
		Au & 1196  &	1060 & 1224 & 1322\\
		\hline
	\end{tabular}}
	
	(a) $\rm T_{50}$s calculated from the GRAINS Code. If $\rm T_{50}$ for an element is less than 300 K, we do not provide it here.
	(b) $\rm T_{50}$s showed in \citet{Lodders:2003}.
	(c) $\rm T_{50}$s used in \citet{Cassen:1996}.
	(d) The condensation temperatures calculated from the GRAINS Code. When the amount of an element in the condensed phase is larger than 2\% of the total amount of the element, we define the corresponding temperature as the condensation temperature.
	Here the pressure is $10^{-4}$ bar.
\end{table}

\section{Model and methods}
\subsection{Algorithm Overview}

\begin{figure}
	\includegraphics[width=0.9\columnwidth]{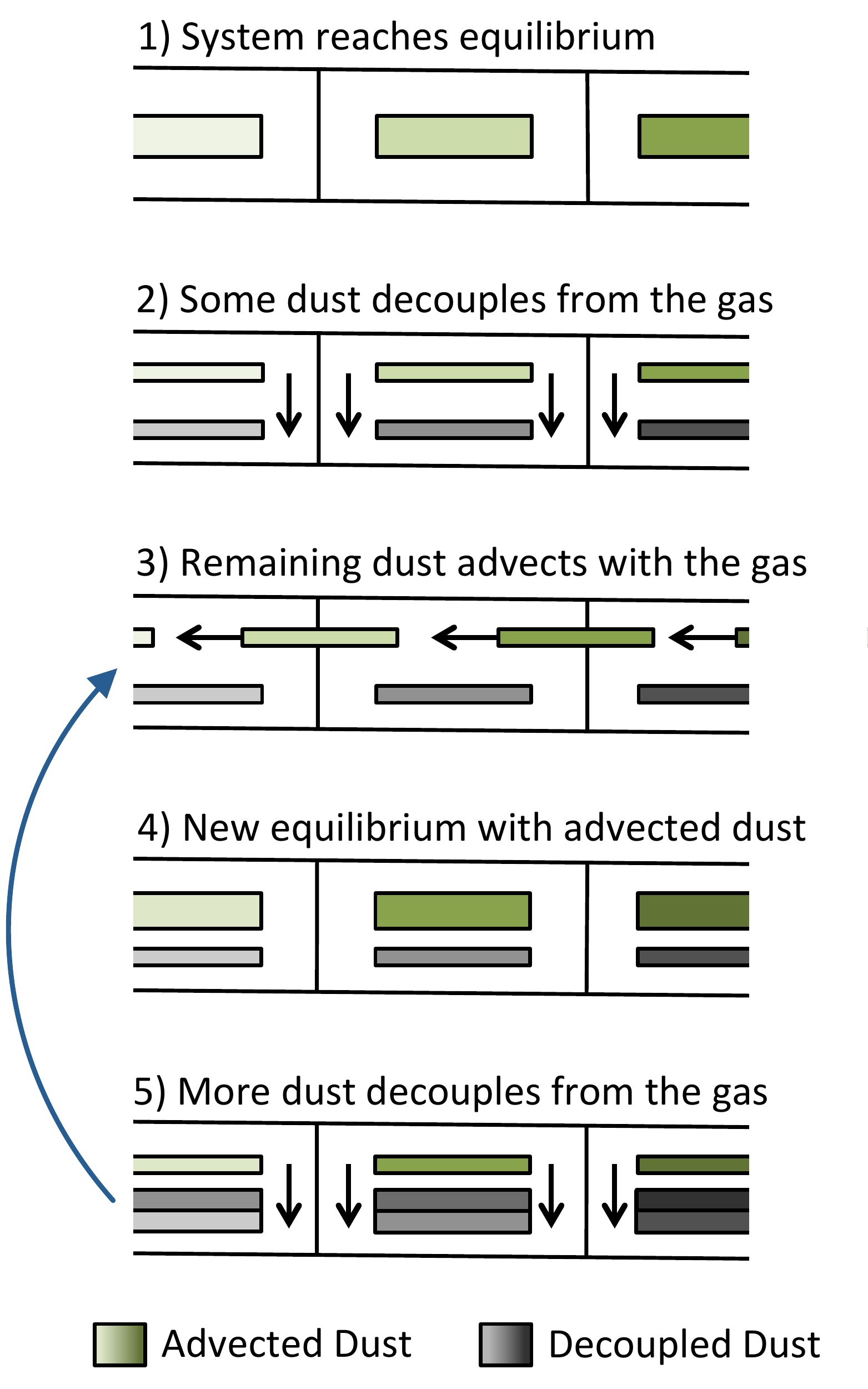}
    \caption{\textbf{Algorithm Flowchart: } A chart showing the calculations done at each step in the evolution of the disc.  We calculate the chemical equilibrium of the gas and advected dust (green) at each radial location.  A portion of the dust then decouples (black) using the decoupling timescale given by Equations (\ref{equ.cond}) and (\ref{equ.tacc}).
    The remaining dust advects to neighboring regions and a new chemical equilibrium is calculated.  The process is repeated as shown by the blue arrow.  Over time the composition of the dust changes as it advects, mixes with the gaseous elements in its new environment, and decouples to settle in the disc midplane.}
    \label{fig:flowchart}
\end{figure}

In our calculations, the elements are partitioned among three phases: one gaseous and two condensed (advected dust and decoupled dust) phases.  The partitioning of elements is based upon the temperature of the disc midplane -- and is either found directly from the full chemical equilibrium calculation, or estimated based on $\rm T_{50}$.  The dust phase decouples from the gas on a timescale proportional to the local orbital period (see Equations (\ref{equ.cond}) and (\ref{equ.tacc})).  The portion of the dust that decouples from the gas remains at the location where it condensed and no longer chemically interacts with other solid or gaseous material.  The portion of the dust that remains coupled to the gas, the advected dust, follows the gas as the disc evolves.
As the advected dust flows into new regions, we assume it is brought into chemical equilibrium with the ambient material and use the new composition to calculate new abundances for the condensed matter at that location.  Figure \ref{fig:flowchart} shows how our algorithm tracks the chemical composition of the decoupled dust in the disc midplane.

\subsection{Gaseous disc evolution model}

For this work, we test our code by adopting the model used by \citet{Cassen:1996} to calculate the evolution of the gaseous disc (H$_2$ and He).  The surface density at semi-major axis $r$ and time $t$ is given by 
\begin{equation}\label{equ.sigma}
\Sigma\left(r,t\right) = \Sigma_{0}\left(t\right)\mathrm{exp}\{-\left[{r}/{r_0}\left(t\right)\right]^2\}.
\end{equation}
Here $ \Sigma_{0}\left(t\right)$ is the surface density near the central star, which is
\begin{equation}\label{equ.sigma0}
\Sigma_{0}\left(t\right) = \frac{M_\mathrm{d}}{\pi {r_0}^2},
\end{equation}
and ${r_0}\left(t\right)$ is the characteristic radius, which is given by
\begin{equation}\label{equ.r0}
{r_0}\left(t\right) = \frac{1}{GM_{\ast}}\left(\frac{J}{M_\mathrm{d}\Gamma \left({5/4}\right)}\right)^2,
\end{equation}
where $M_\mathrm{d}$ is the mass of the disc, $G$ is the gravitational constant, $M_{\ast}$ is the mass of the central star (in this paper, we use 1 $M_{\odot}$), $J$ is the angular momentum of the disc (we adopt $J=3\times10^{52}\ \mathrm{g\ cm^2\ s^{-1}}$), and $\Gamma$ is the Gamma function (i.e.,  $\Gamma\left({5/4}\right)=0.9064$).

The evolution of the disc mass is governed by
\begin{equation}\label{equ.md}
{M_\mathrm{d}}\left(t\right) = {M_\mathrm{d0}}\left(1+\frac{t}{t_\mathrm{e}}\right)^{-0.5},
\end{equation}
where ${M_\mathrm{d0}}$ ($0.21 \ \rm M_{\odot}$) is the mass of the disc at $t=0$, and $t_\mathrm{e}=2.625\times10^4$ yr is the characteristic evolution timescale of the disc.

The midplane of the disc is heated by the accretion of the gas, so the midplane temperature, $T\left(r,t\right)$, can be calculated from \citep{Cassen:1994}
\begin{equation}\label{equ.tem}
T^4 = \frac{3 G \tau M_{\ast} \dot{M}_{\rm I}}{64 \pi \sigma_{\rm SB} r^3},
\end{equation}
where $\dot{M}_{\rm I}$ is the mass accretion rate to the central star, $\sigma_{\rm SB}$ is the Stefan-Boltzmann constant, and $\tau={\kappa\Sigma}/{2}$ is the optical depth ($\kappa=4\ \mathrm{cm}^2\ \mathrm{g}^{-1}$ is the opacity of gas with Solar composition).  For simplicity, we use a constant opacity throughout the disc.

The midplane pressure is
\begin{equation}\label{equ.p}
P = \frac{\rho \mathcal{R} T}{\mu},
\end{equation}
where $\mathcal{R}$ is the gas constant, $\mu=2.34$ is the mean molecular weight in atomic mass units, $\rho=\Sigma/\sqrt{2\pi}H$ is the midplane density, and $H=\sqrt{\mathcal{R}T/\mu}/\Omega$ is the scale height of the gaseous disc.

We note that our disc is not in a steady-state.  The mass accretion rate of the gas is given by
\begin{equation}\label{equ.vr}
\dot{M}\left(r\right) = -1.5\Sigma\nu \left(1-4\left(\frac{r}{r_0}\right)^2\right).
\end{equation}
Over time, the value of $r_0$ increases from 3.2 AU at the beginning to 19 AU at $5t_{\rm e}$.  The radius at which the accretion rate equals zero moves outwards, and the behavior of the disc in the inner region is increasingly similar to a steady state disc.  However, we overestimate the temperature in the inner region at early evolution of the disc, which affects slightly the location where the elements begin to condense and where the composition of the decoupled materials match that in terrestrial planets and chondrites.

\begin{figure}
	\includegraphics[width=\columnwidth]{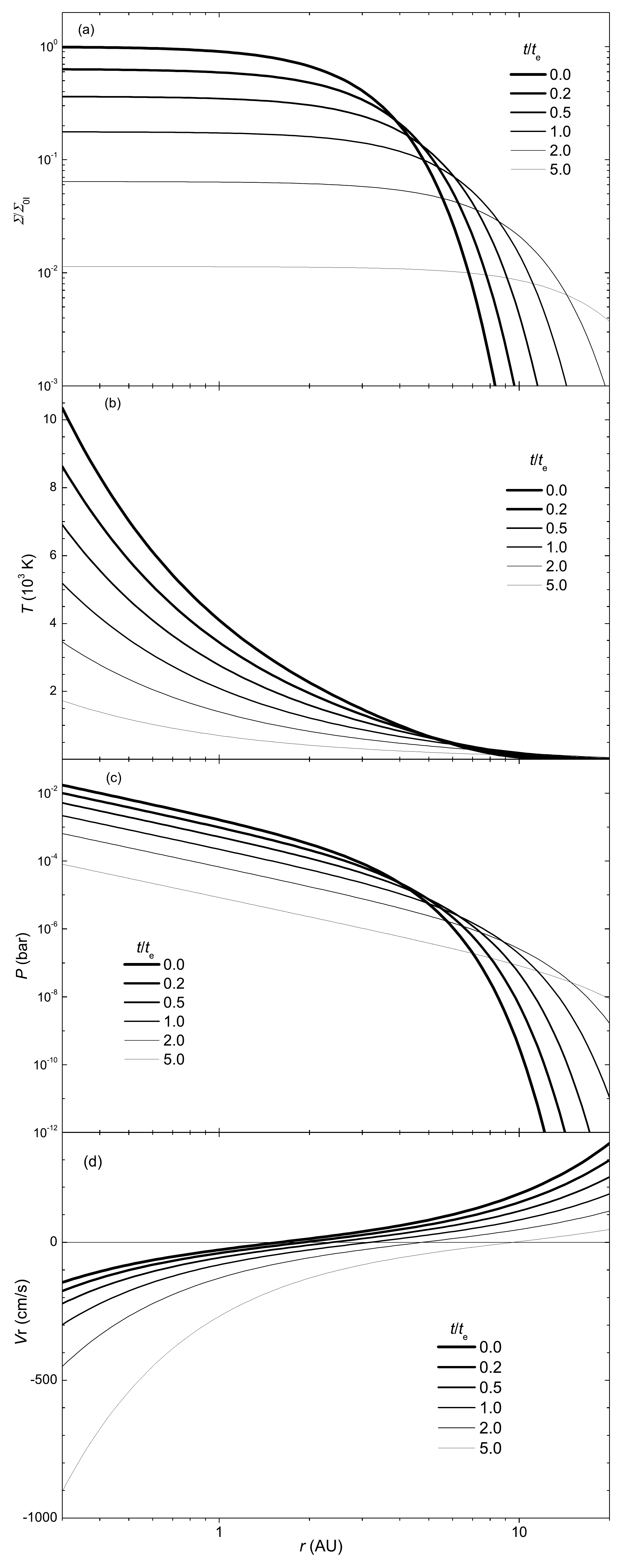}
    \caption{\textbf{Disc evolution: } (a) Normalized surface density ($\Sigma_{\rm 0I}$ is the initial value of $\Sigma_{0}$), 
    (b) Midplane temperature, and (c) Midplane pressure for the disc at various points in time, (d) Radial velocity of the gas.}
    \label{fig:sigma}
\end{figure}

Figure \ref{fig:sigma} shows how the disc evolves with time. Figure \ref{fig:sigma} (a) shows the evolution of the normalized surface density of the disc.  As time goes on, the surface density decreases in the inner region (radii less than several AU) while it increases in the outer region of the disc (i.e., the disc expands).  The general trends of temperature (Fig. \ref{fig:sigma} (b)) and pressure (Fig. \ref{fig:sigma} (c)) with time are similar to the trend of surface density.  Note that within 4AU, the temperature decreases with time.

\subsection{Motion and decoupling of condensable compounds}

We assume that all elements in the gas phase have the same velocity as the H$_2$ and He \citep{Jacquet:2012}.  Their evolution is governed by
\begin{equation}\label{equ.sigmat}
\frac{\partial\Sigma_i}{\partial t}+\frac{1}{r}\frac{\partial\left(r V_r \Sigma_i\right)}{\partial r}=-F_{i},
\end{equation}
where $\Sigma_i$ is the surface density of element $i$ in the gaseous and advected dust phases, $V_r$ is the radial velocity of H$_2$ and He, and $F_{i}$ is the depletion rate due to the condensation. In this work, we only consider the advection of particles.  We do not consider diffusion from turbulence, which may play a role in turbulent discs.

In our models, once an element begins to condense it also begins to decouple from the gas phase.  The amount of the decoupled dust is determined by a decoupling rate that depends upon the surface density of the advected dust,
\begin{equation}\label{equ.cond}
\frac{{\rm d}\sigma_{i}}{{\rm d} t}=F_{i} =\epsilon \Omega \Sigma_{i,\rm adv},
\end{equation}
where $\sigma_i$ is the surface density of element $i$ in the decoupled dust,
$\epsilon$ is the decoupling efficiency, $\Omega$ is the angular velocity around the central star, and $\Sigma_{i,\rm adv}$ is surface density of element $i$ in the advected dust. 
For the method that uses $\rm T_{50}$, $\Sigma_{i,\rm adv}$ is a step function around $\rm T_{50}$ from 0 to $\Sigma_i$.  
For the chemical equilibrium method
$\Sigma_{i,\rm adv}$ is determined by chemical equilibrium calculation of the system. 
Initially, all the elements are of Solar abundance.  After getting $\Sigma_{i,\rm adv}$ and $V_r$ at each time and each radius, we integrate $\Sigma_i$ using the Euler method.

We assume the decoupling efficiency is uniform across the entire disc.  The decoupling timescale at 1 AU is given by
\begin{equation}\label{equ.tacc}
t_{\rm dec}=\frac{1}{\epsilon \Omega \left(r=1 \rm AU\right)}.
\end{equation}
At a given radial distance, the time for the dust to decouple depends upon the local orbital speed and we normalize the decoupling time to the orbital speed at 1 AU.  A decoupling timescale of zero corresponds to immediate decoupling, while an infinite timescale implies no decoupling at all.  Generally, the timescale for dust to coagulate into planetesimals is $10^3$ to $10^5$ years \citep{Weidenschilling:1993}.  For our fiducial models, we adopt a disc evolution timescale of $t_\mathrm{e}=2.625\times10^4$ yr and a dust decoupling timescale of $t_{\rm dec}=1.5\times10^4$ yr, which is the same as that used in \citet{Cassen:1996}.  All these models, and the parameters we used for them are shown in Table \ref{tab:para} where the historical model \citet{Cassen:1996} is labeled as MC, and the two new models are labeled as M1 and M2 (described below).

\begin{table*}
	\centering
	\caption{Different models and disc evolution and dust decoupling timescales.}
	\label{tab:para}
	\begin{tabular}{l|c|c|c|c|c|c|c}
		\hline\hline
		 Timescales &\multicolumn{1}{c}{Historical model} & \multicolumn{2}{c}{Models with new $\rm T_{50}$} & \multicolumn{4}{c}{Time-dependent chemical equilibrium model} \\
		 &\multicolumn{1}{c} {(MC)} & \multicolumn{2}{c} {(M1)} & \multicolumn{4}{c}{ (M2)} \\
		\hline
		  & Typical values & Typical values& Group 3 & Typical values & Group 2 & Group 3 & Group 4 \\
		   & (MC)  & (M1G1) & (M1G3) & (M2G1)  & (M2G2) & (M2G3) & (M2G4) \\
		\hline
	     $t_{\rm dec}$/yr & 1.5e4 & 1.5e4 &1.5e3& 1.5e4 & 1.5e5 & 1.5e3 & 1.5e6\\
        \hline
		 $t_{\rm e}$/yr & 2.625e4 & 2.625e4& 2.625e4 & 2.625e4 & 2.625e4 & 2.625e4 & 2.625e6\\
		\hline
	\end{tabular}
\end{table*}

\subsection{Model 1 and recalculating the 50\% condensation temperatures with GRAINS}\label{sec:grain}

In our first model (M1), an element begins to condense out of the gas once the midplane temperature drops below $\rm T_{50}$ (as was done in \citet{Cassen:1996}).  In order to compare our results with previous work, we first recalculate $\rm T_{50}$ using the GRAINS code \citep{Petaev:2009}.  
The elements used in GRAINS are listed in Table \ref{tab:element}.  In this case, H and He account for 98.65\% (H: 71.34\%, He: 27.31\%) of the total mass with the remaining elements having Solar abundances \citep{Lodders:2003}.  GRAINS distributes 33 elements among 242 gaseous and 520 solid species by minimizing the Gibbs free energy of the system to identify the chemical equilibrium composition.

\citet{Cassen:1996} and \citet{Lodders:2003} estimate their $\rm T_{50}$s based upon a single pressure of $10^{-4}$ bar.  However, the $\rm T_{50}$ depends upon both the pressure and the starting composition, which, in turn, affects the condensation process throughout the disc.  Here, we recalculate $\rm T_{50}$ for each element by fixing the value of pressure (from $10^{-12}$ to $1$ bar) and lowering the temperature from 2500 K to 300 K.  We note that this calculation does not involve the disc evolution---so there are only the gas and dust phases. 

\begin{figure}
	\includegraphics[width=0.95\columnwidth]{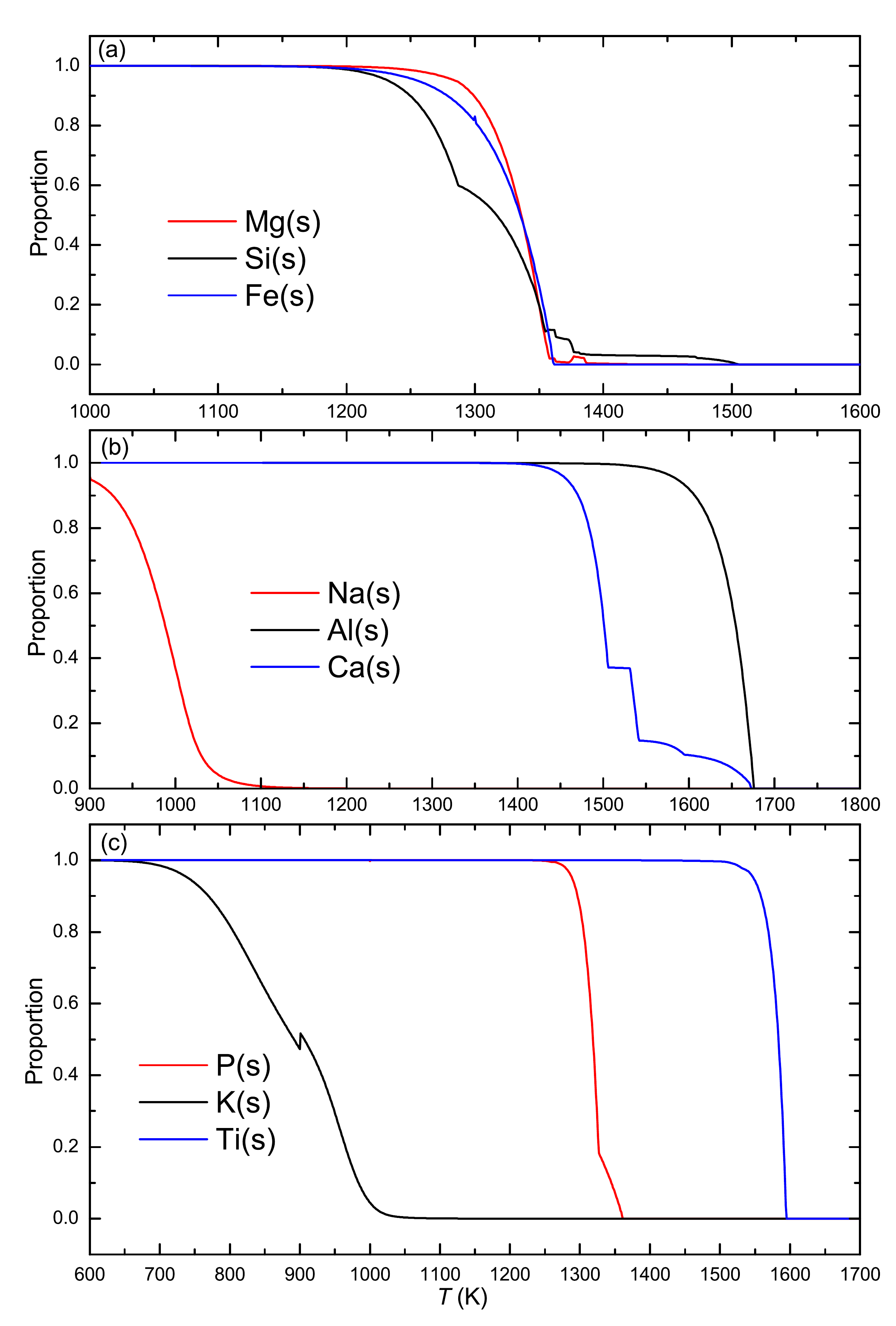}
    \caption{\textbf{Condensation curves: }
    Proportions of elements (Na, Mg, Al, Si, P, K, Ca, Ti, and Fe) in the condensed phases as a function of the midplane temperature.  The relative abundances of each element match the solar composition and the pressure used in these calculations is $10^{-4}$ bar.
    }
    \label{fig:con-tem0}
\end{figure}

The condensation histories of several elements as functions of temperature are shown in Figure \ref{fig:con-tem0}.  In Figure \ref{fig:con-tem0} (a), we see that $\rm T_{50}$ for Mg, Si, and Fe are 1336, 1316, and 1335 K, respectively.  Due to the difference between the temperatures where an element begins to condense and where it is completely condensed, the element with lower $\rm T_{50}$ may have higher relative abundance in the dust phase at certain temperatures. 

\begin{figure}
	\includegraphics[width=\columnwidth]{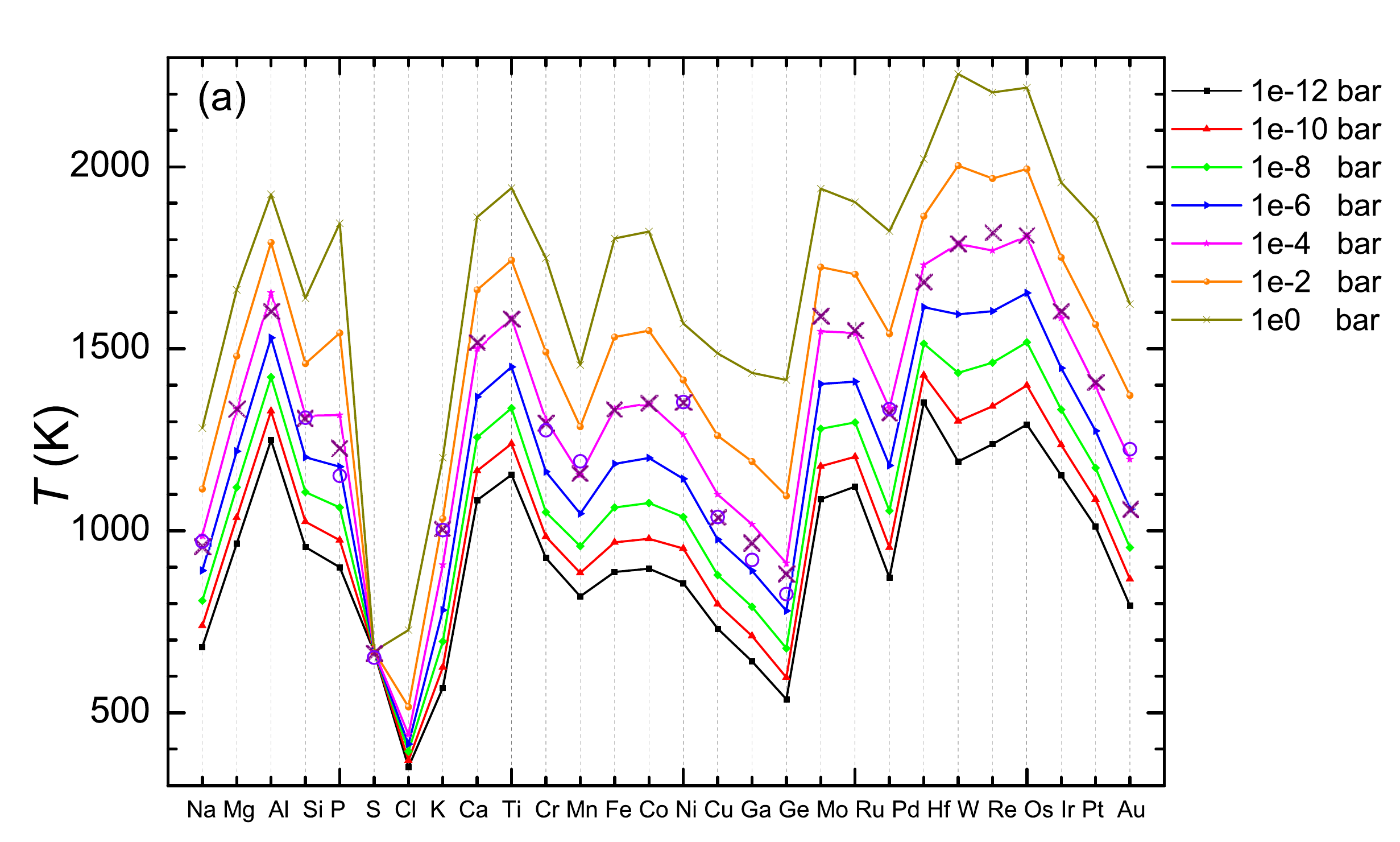}
	\includegraphics[width=\columnwidth]{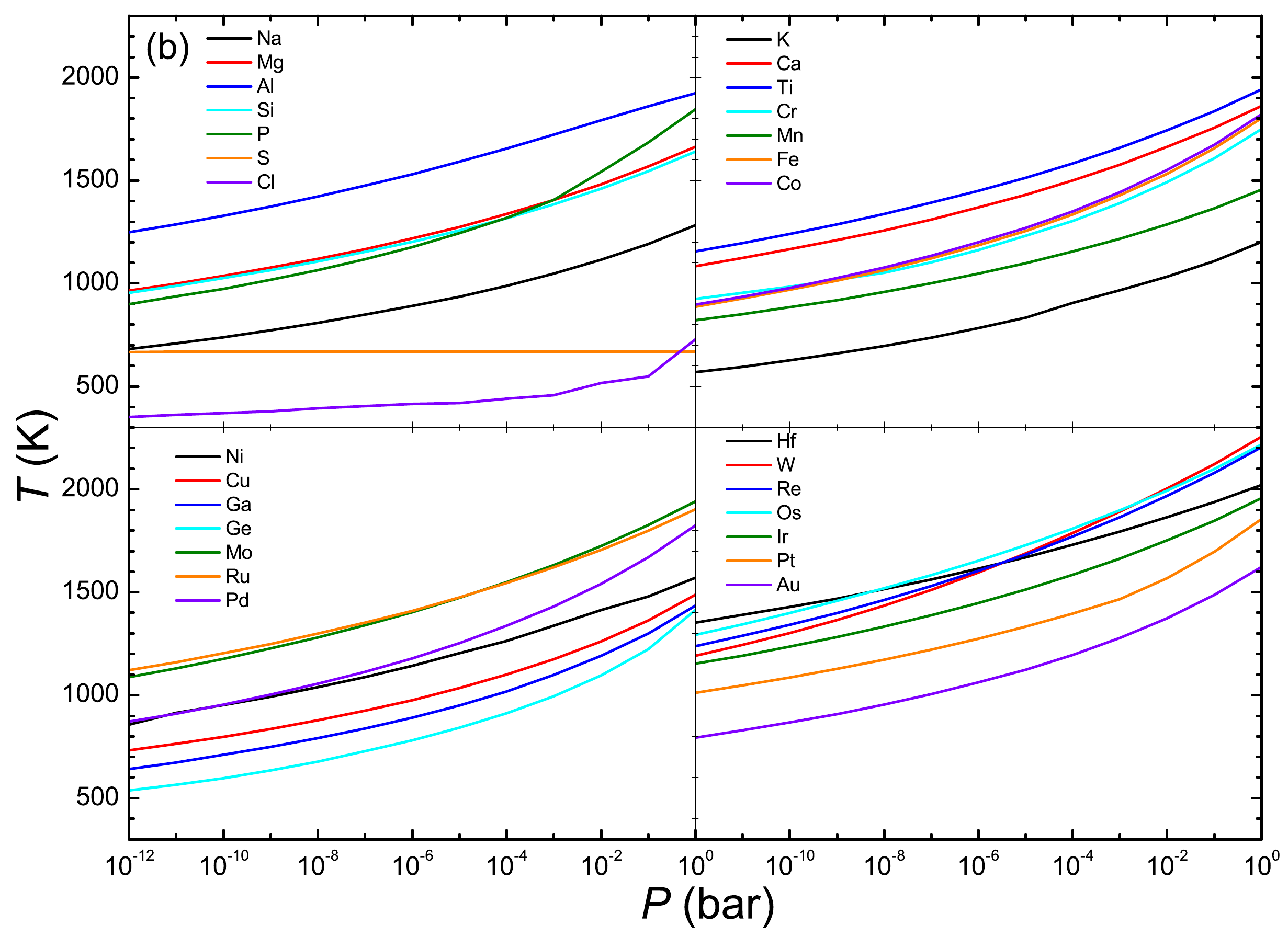}
    \caption{\textbf{50\% condensation temperatures: } $\rm T_{50}$ as a function of elements for various pressures (a) and pressure for various elements (b).  In panel (a), we also add $\rm T_{50}$ used in \citet{Cassen:1996} (violet circle) and \citet{Lodders:2003} (purple cross) with pressure $10^{-4}$ bar and Solar composition.}
    \label{fig:t50-pressure}
\end{figure}

Figure \ref{fig:t50-pressure} shows the effect on the $\rm T_{50}$s when changing the pressure from $10^{-12}$ to 1 bar.  When the pressure increases, the $\rm T_{50}$s also increases, and the relative values of $\rm T_{50}$ may change.  
Also shown in that figure are the $\rm T_{50}$s used in \citet{Cassen:1996} and in \citet{Lodders:2003} with pressure $10^{-4}$ bar and Solar composition.   While the values are generally consistent with each other, several elements have significant differences (e.g., K, Cu, Ni, and P). 
\citet{Woitke:2018} calculated chemical equilibrium for solar abundance at 1 bar.  The $\rm T_{50}$s at 1 bar here are consistent with the values from Figure 5 in their paper.
With the new $\rm T_{50}$s shown above, we use Equation (\ref{equ.cond}) to calculate the decoupling of the elements as in \citet{Cassen:1996}.

\subsection{Model 2: Time-dependent chemical equilibrium model}

In our second model (M2), we use the GRAINS code to calculate the chemical equilibrium dynamically throughout the evolution of the disc rather than assuming an immediate condensation at $\rm T_{50}$ followed by the decoupling process.  In this model, $\Sigma_{i, \rm adv}$ is calculated from the chemical equilibrium of the system, and grows smoothly as the temperature falls below the point where the elements begin to condense (see Table \ref{tab:element} and Figure \ref{fig:con-tem0}).  
Note that the temperature where an element begins to condense (T$\rm _2$, in Table 1) is higher than $\rm T_{50}$ (where half of the element has already condensed, see Table \ref{tab:element}).

\begin{figure}
	\includegraphics[width=0.95\columnwidth]{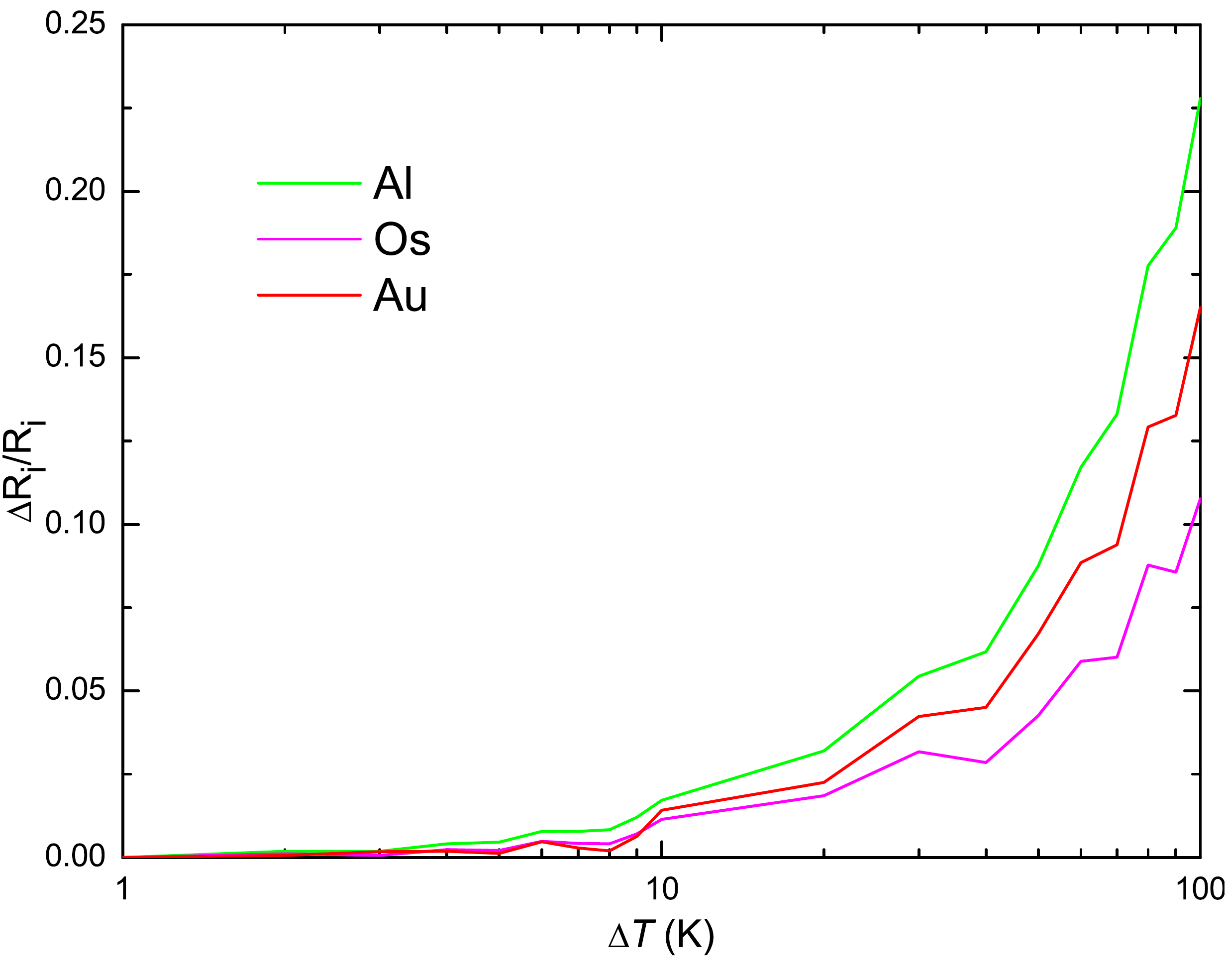}
    \caption{\textbf{Relative values of $\rm R_i$ for different temperature steps: } Relative values of $\rm R_i$, for a sample of elements, as a function of the change in temperature that triggers the evaluation of the chemical equilibrium.  $\Delta \rm Ri$ is the difference between $\rm R_i$ for temperature steps $\Delta T$ and that for 1 K. $\rm R_i$ in y-axis is the value for 1 K. Evaluating the chemical equilibrium when the temperature at a distance of 1.8 AU drops by 10K produces differences at the level of 1\%.  The abundance differences shown in this plot are evaluated at 3 AU.
    }
    \label{fig:delt}
\end{figure}

To save computational time, we do not determine the chemical equilibrium at every timestep of the disc evolution.  Rather, we calculate the chemical equilibrium everywhere in the disc whenever the midplane temperature at 1.8 AU changes by 1 K.  These temperature steps at this radius are small enough to keep the size of the temperature steps at all other relevant radii small enough that the chemical equilibrium calculation is minimally affected.  We verified that these choices do not affect our results by testing the convergence of the condensation results with several simulations where we fix the temperature step at 1.8 AU to be several values in the range between 1 and 100 K.  Figure \ref{fig:delt} shows that if the temperature step is less than or equal to 10 K, the differences between the results are less than 1\%.  Consequently, we need not use a temperature step smaller than 1 K -- which yields errors at a level near $10^{-3}$.

\section{Results}

Our first new model (M1) produces results which differ only slightly from the model from \citet{Cassen:1996} (MC).  Thus, we do not discuss the details of that model.  Rather, we first discuss the evolution of the decoupled dust from the second model after which we compare the results of the two models (M1 and M2).

\subsection{Evolution of the surface density of decoupled dust}

\begin{figure}
	\includegraphics[width=\columnwidth]{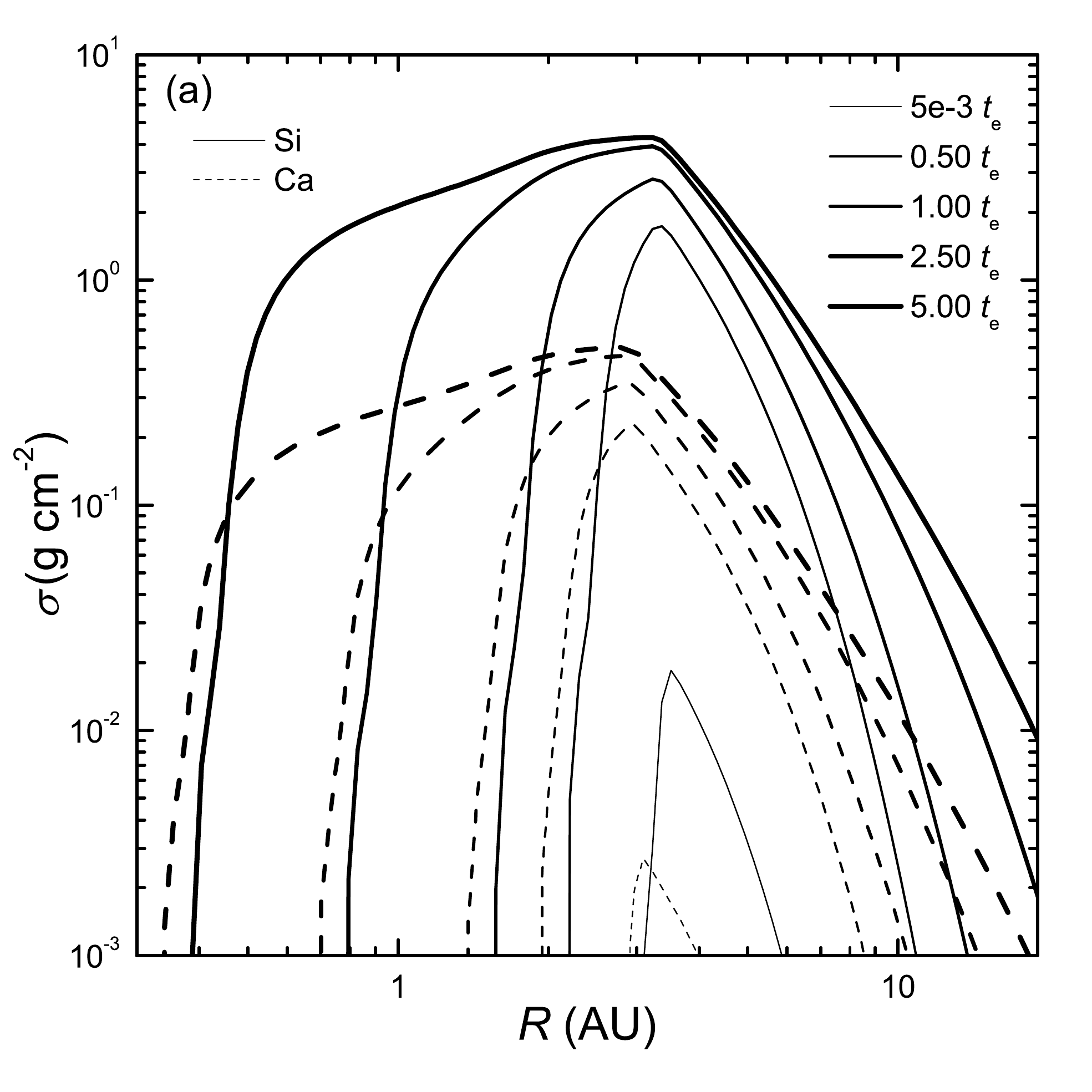}
    \caption{\textbf{Condensed surface density and relative abundance over time: }
        (a) Evolution of Si and Ca in the decoupled dust for the model (M2G1).  Solid lines denote the evolution of Si while dashed lines indicate Ca.  Over time the condensation fronts move inward as the temperature decreases and outward as the surface density rises.  
 }
    \label{fig:sigmaevo}
\end{figure}

To give an example of how a system evolves, Figure \ref{fig:sigmaevo} (a) shows how Si and Ca in the decoupled dust evolve with time. Si first condenses at around 4 AU.  At smaller radii, the temperature of the disc is too high for Si to condense, while at larger radii the gas surface density is low -- yielding a surface density for the decoupled materials that is similarly low.  As time goes on, the condensation region expands both inward (as the temperature of the inner region decreases) and outward (as the gas surface density increases due to viscosity and the outward flow of disc material).

The general trend of the evolution of Ca is similar to that of Si except the maximum surface density and the inner and outer boundaries of the condensed region change.  The different maximum surface densities and outer boundaries are determined by the different gas surface densities of the elements and their condensation temperatures.  The inner boundaries are correlated with the $\rm T_{50}$ for the elements.

\subsection{Comparison of the two models}

\begin{figure}
	\includegraphics[width=\columnwidth]{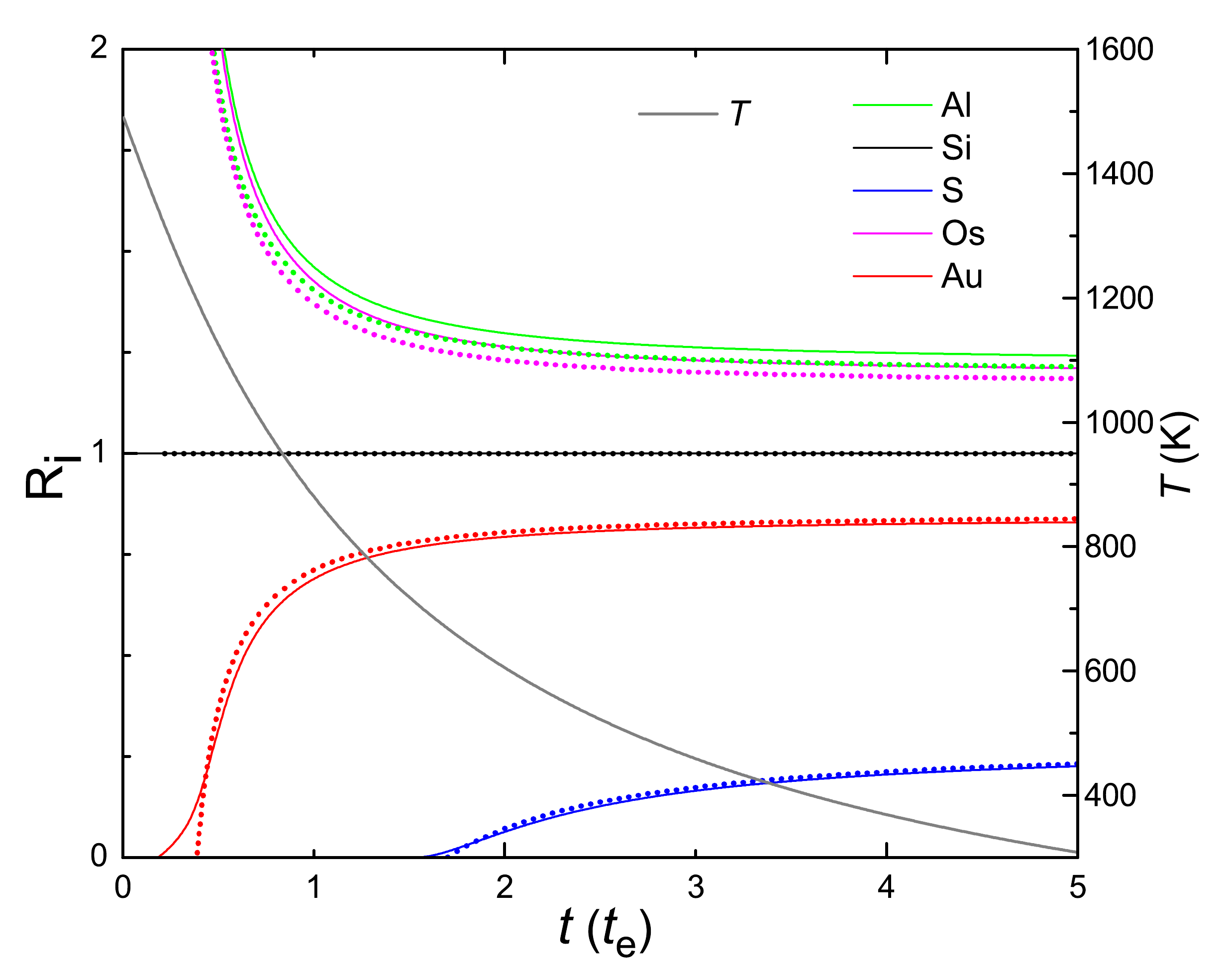}
    \caption{\textbf{Abundances vs time: } $\rm R_i$ as a function of time for Al, Si, S, Os, and Au at 3 AU.  Al and Os are refractory elements, which begin condensing early while S and Au are more volatile and condense later.  Both chemical equilibrium calculation (M2G1, solid lines) and fixed $\rm T_{50}$ (M1G1, dotted lines) of the elements in the evolving disc are plotted (the models used in  
    \citet{Cassen:1996}) and in Figure \ref{fig:1e4compare}).  The ratios of the final values of $\rm R_i$ between the two models for Al, S, Os, and Au are 0.979, 1.03, 0.978, and 1.01, respectively with refractory elements being underestimated and volatile elements being overestimated by the 50\% condensation temperature approximation.  The corresponding temperature (gray line) is also plotted with the temperature scale along the right side.
}
    \label{fig:rit-compare}
\end{figure}

In Figure \ref{fig:rit-compare}, we show the evolution of $\rm R_i$ for Al, Si, S, Os, and Au at 3 AU for our two models M1 and M2.  Al and Os are both refractory elements with $\rm T_{50}$ for Al (1654 K) and Os (1812 K) both higher than that for Si (1316 K).  Thus, their $\rm R_i$ decrease over time as more Si condenses.  The final ratios calculated from the chemical equilibrium are slightly larger than those calculated using $\rm T_{50}$.  Both S and Au are more volatile with their $\rm T_{50}$s being lower than that for Si: S (669 K) and Au (1196 K).  Thus, their $\rm R_i$ increases over time from 0 to their final values.

\begin{figure}
	\includegraphics[width=\columnwidth]{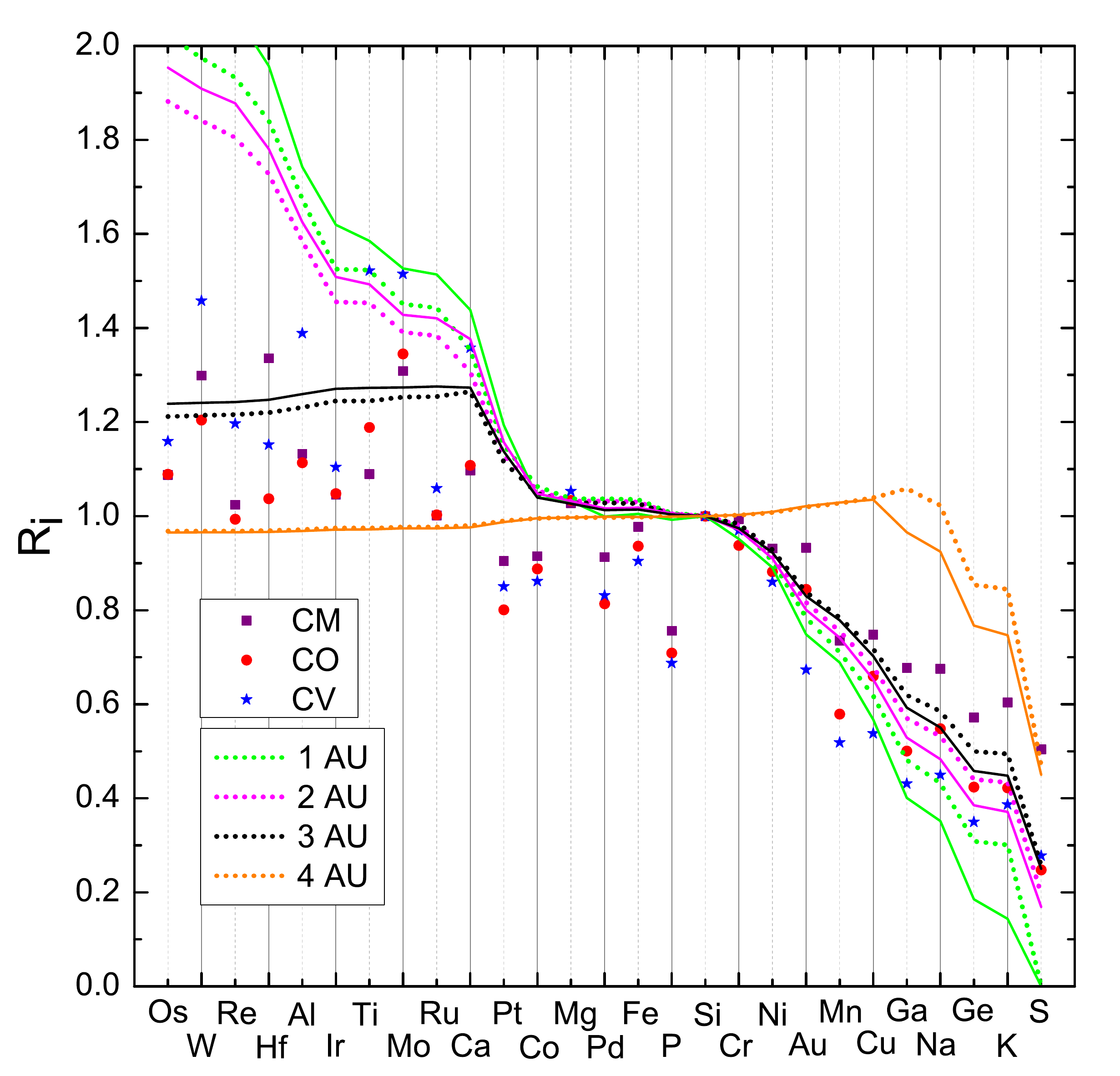}
    \caption{\textbf{Equilibrium vs 50\% condensation models (M1G1 vs M2G1): } Comparison of $\rm R_i$ calculated from the two new models at the end of the simulation ($5t_{\rm e}$, where $t_{\rm e}=2.625\times 10^4$ yr).  The chemical equilibrium model (M2G1, solid lines) and the fixed $\rm T_{50}$ model (M1G1, short dotted lines) for the elements in the evolving disc are shown.}
    \label{fig:1e4compare}
\end{figure}

Since the dust begins to condense earlier in the chemical equilibrium calculation (M2 models) than it does in the $\rm T_{50}$ calculations (M1 models) and because the surface density of the disc decrease quickly at early time in the inner region, the approximation from $\rm T_{50}$ systematically overestimates the volatile abundance (3\% for S and 1\% for Au) while it systematically underestimates the refractory abundances (2\% each for Al and Os).  The dotted lines in Figure \ref{fig:1e4compare} show $\rm R_i$ by using updated $\rm T_{50}$s calculated from the GRAINS code (M1G1).  There are more refractory elements in Figure \ref{fig:1e4compare} than 
are shown in the results of \citet{Cassen:1996} because the GRAINS Code considers more elements.

The general trends at 1 and 2 AU are similar to those from the historical model.  At 3 AU, $\rm R_i$ for elements with $\rm T_{50}$ higher than Ca (1500 K) are almost the same, because the initial temperature at this radius is near $\rm T_{50}$ for Ca and these elements begin condensing simultaneously.  Because the temperature interior to 4 AU decreases with time, the value of $\rm R_i$ for an element depends upon the initial value of the temperature at a given radius.  If the initial temperature at a point is lower than $\rm T_{50}$ for Si (1316 K), $\rm R_i$ will be 1.  If it is higher than $\rm T_{50}$, $\rm R_i$ will decrease with decreasing $\rm T_{50}$.  At 4 AU, the $\rm R_i$ of the two models are largely the same for elements with $\rm T_{50}$ higher than that for Cu (1100 K).

\section{Quantifying the effects of decoupling and disc evolution timescales}

\begin{figure}
	\includegraphics[width=\columnwidth]{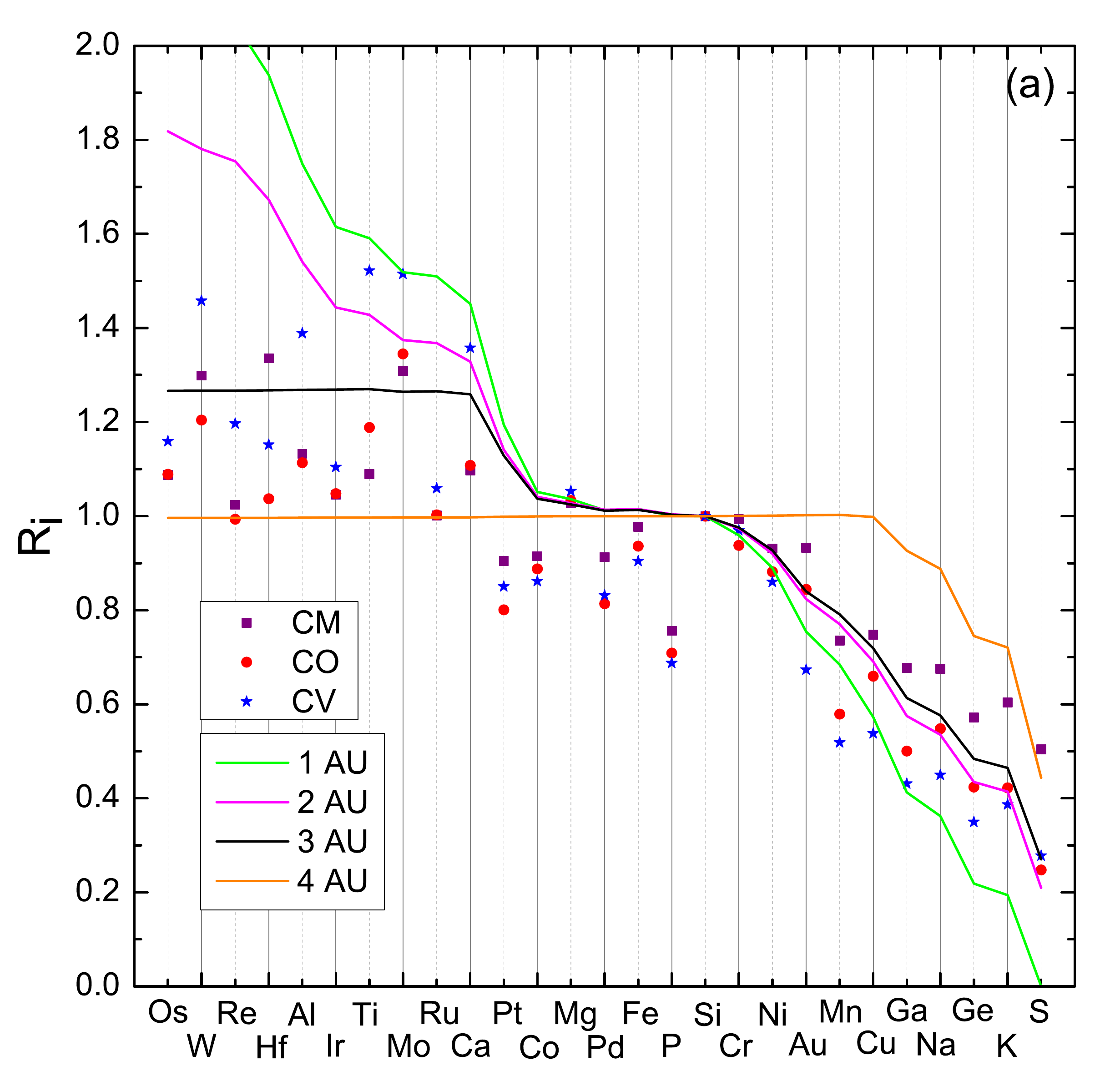}
	\includegraphics[width=\columnwidth]{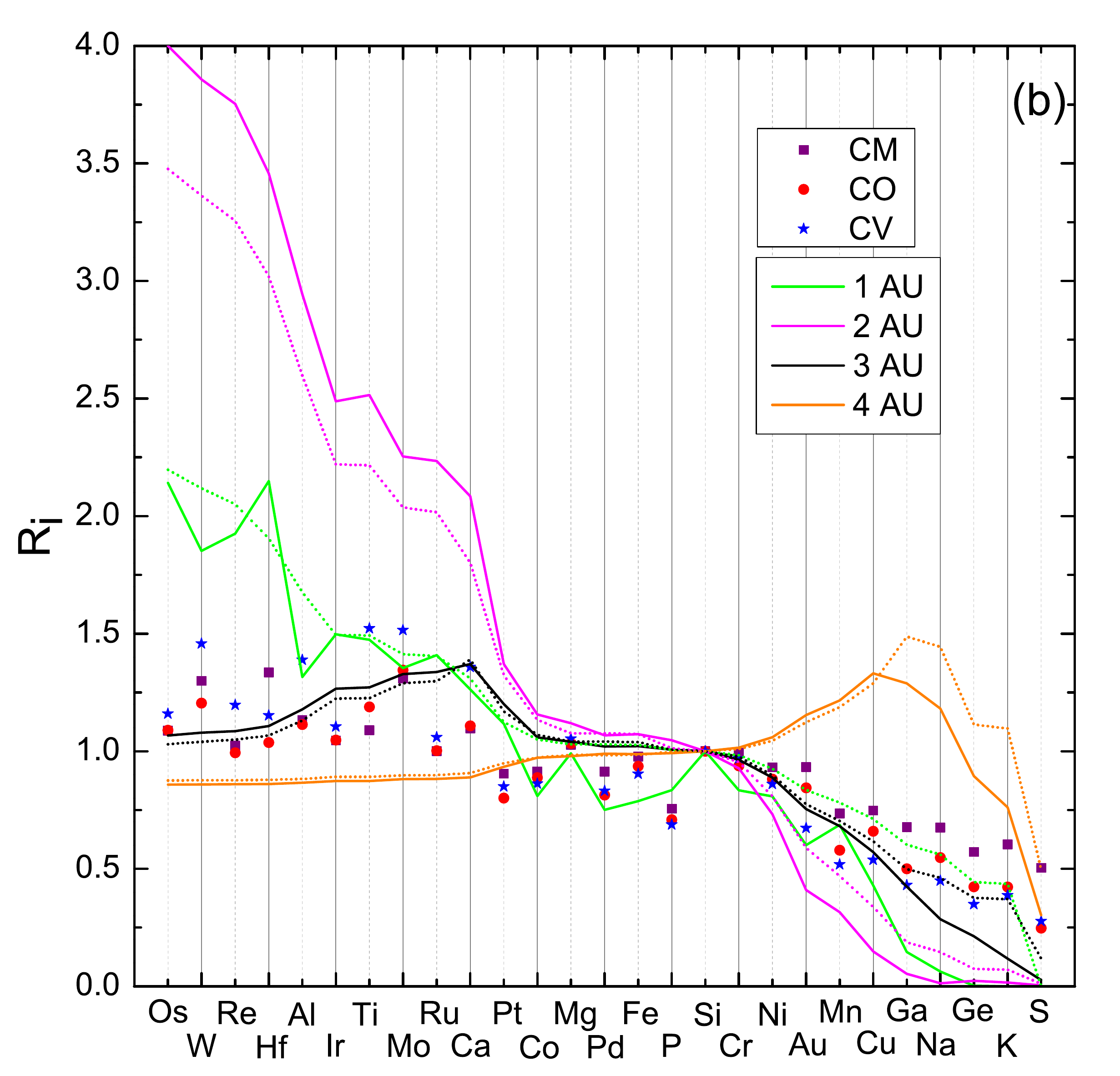}
    \caption{\textbf{Effects of changes to the decoupling timescale (M2G2 and M2G3): }  These panels show the effect of changing the decoupling timescale of the dust relative to the disc evolution timescale.  
    Values of $\rm R_i$ at different radii and at time $t=5\ t\rm{_e}$ ($t\rm{_e}=2.625\times10^4\ yr$)
    calculated from the evolution of disc and using the GRAINS code.  The decoupling timescales here are (a) $1.5\times10^5$ yr (M2G2) and (b) $1.5\times10^3$ yr (M2G3), which are of one order of magnitude longer or shorter than that in the typical case respectively.  
    The elements are ordered from high to low $\rm T_{50}$ calculated from the GRAINS code.
    In panel (b), we also plot the results calculated from $\rm T_{50}$s (M1G3, dotted lines).  
    Note that with the longer decoupling time there is a consistent trend in the relative abundances with distance.  With the short decoupling time we observe peaks in the abundances where the initial temperature in the disc correlates with the condensation temperature of the element at that particular radial distance.  Also, with short condensation times, the innermost region (where the gas flows toward the star rather than away) the refractory elements are depleted somewhat as the incoming gas from more distant regions contains a higher proportion of volatile elements (see text for details).
    }
    \label{fig:cassen-misha-final1e53}
\end{figure}

\subsection{Changes to the decoupling timescale}

To illustrate the conditions under which our simulations can reproduce the chondrite abundances, we look at the effects of changes to the disc evolution timescale and decoupling timescale.  We first change the decoupling timescale, $t_{\rm dec}$, to be an order-of-magnitude longer than the nominal case ($1.5\times 10^5$ yr -- M2G2, Figure \ref{fig:cassen-misha-final1e53} (a)).  We find that these results are almost the same as those for the nominal case.   Because the decoupling timescale is very long, the material flows with the gas and leaves behind the same relative abundances of decoupled dust.  Small $\epsilon$ does not affect the relative $\Sigma_{i,\rm adv}$ and, consequently, not the $\sigma_i$.  Thus, while less material decouples overall, the relative amount of the elements in the decoupled dust stays almost the same -- a result that persists with even longer decoupling timescales.

Figure \ref{fig:cassen-misha-final1e53} (b) shows the corresponding abundances at different radii when the decoupling timescale is an order-of-magnitude shorter than the nominal timescale (M2G3, $1.5\times 10^3$ yr).  
Compared to Figure \ref{fig:1e4compare}, the abundances at 3 AU peak at Ca, with a $\rm T_{50}$ of 1500 K.  The more refractory elements have lower $\rm R_i$, because they condense and quickly decouple in the inner regions and do not flow outward.  The most refractory elements at 3 AU have the lowest $\rm R_i$.  Ca has the largest abundance because it condenses and decouples at the beginning of the simulation and additional Ca is advected into that region from the inner disc (where it was too hot for Ca to condense) while the more refractory elements remain behind.  At 4 AU, a similar effect happens with Cu. The more refractory elements also have lower $\rm R_i$ while the more volatile elements have higher $\rm R_i$ for elements with $\rm T_{50}$ larger than that for Ca (1100 K).

The overall effect of the relationship between the disc evolution timescale and the dust decoupling timescale (i.e., the planetesimal formation timescale) is that there will be peaks in the radial distribution of different elements that depend upon their condensation temperatures.  If planetesimals form quickly, then the more refractory elements will not flow with the gas, and regions fed by the warmer parts of the disc will be somewhat depleted in those elements.  On the other hand, if planetesimals form more slowly (a long decoupling timescale), then the advecting refractory materials will enrich neighboring parts of the disc.  Definitive statements about the consequences of this effect will require a more realistic disc model and lie beyond the scope of this work.

\subsection{Changes to the disc evolution timescale}

To see how the disc evolution timescale affects $\rm R_i$, we change both the decoupling timescale and the disc evolution timescale to be two orders-of-magnitude longer than the typical case while keeping the ratio the same (M2G4).  With the disc evolution timescale two orders-of-magnitude longer than the typical case, the accretion rate from the disc to the central star is smaller by two orders-of-magnitude, and the temperature is approximately one third of the typical case (See Equation \ref{equ.tem}).

\begin{figure}
	\includegraphics[width=\columnwidth]{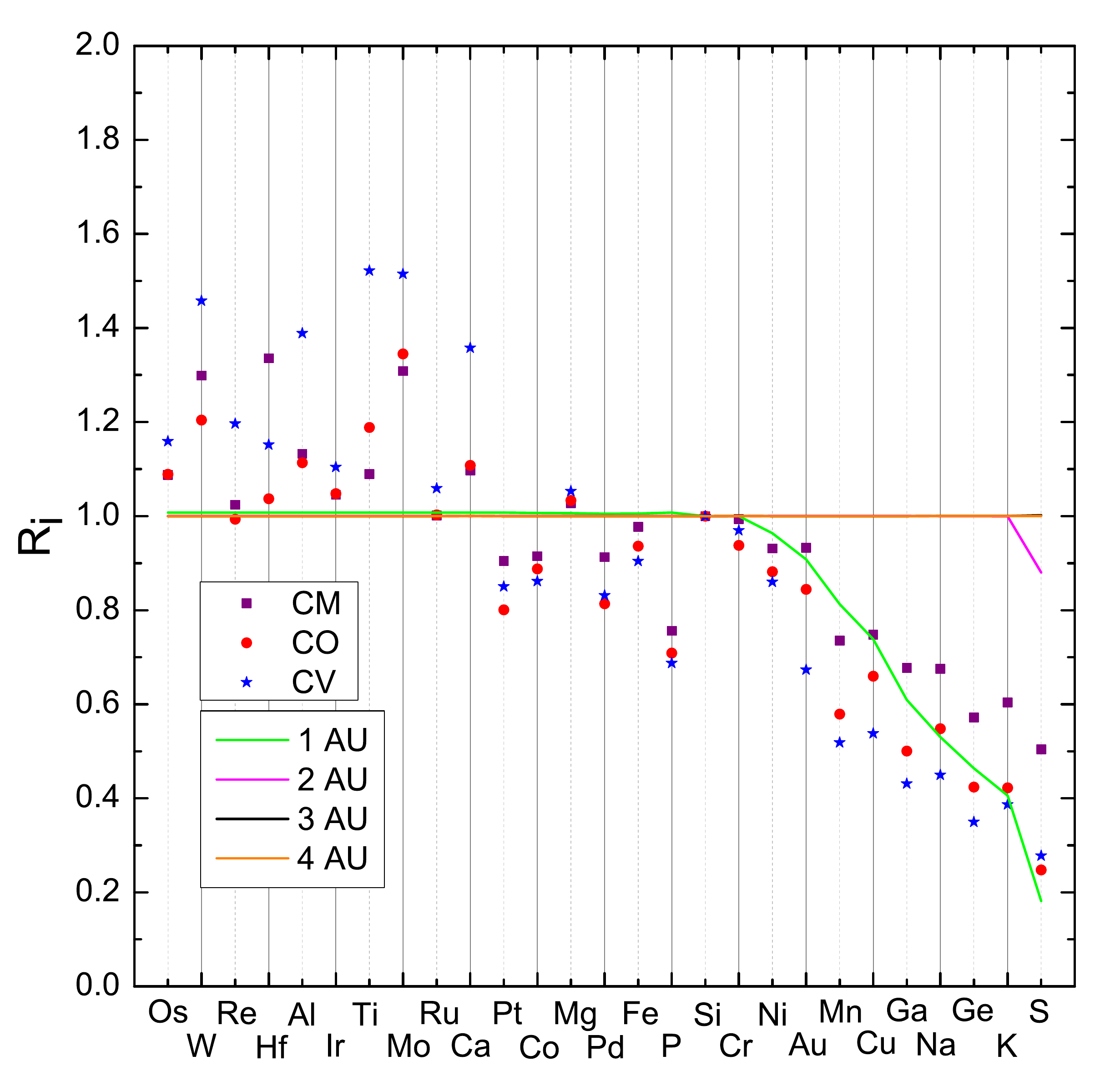}
    \caption{\textbf{Changing both disc and dust evolution timescales (M2G4): } 
    Values of $\rm R_i$ at different radii calculated from the evolution of disc and condensation from the GRAINS Code.  The decoupling timescale and disc evolution time scales are two orders-of-magnitude longer than the typical case ($t\rm_{dec} = 1.5\times10^6$ yr and $t\rm{_e}=2.625\times10^6$ yr respectively).  The lower temperature of the disc, except at the innermost location, causes most elements to condense simultaneously with Si.  The elements are ordered from high to low $\rm T_{50}$ calculated from the GRAINS code.
}
    \label{fig:cassen-misha-final1e66}
\end{figure}

Because the temperature across the disc is now much lower, and is even below $\rm T_{50}$ of Si at 1 AU (1316 K), most of the elements condense immediately at the start of the simulation. 
The most volatile elements that are close to the central star condense later than the refractory elements and have lower $\rm R_i$.   
The resulting abundances are shown in Figure \ref{fig:cassen-misha-final1e66}.  The refractory elements at different radius all have the same $\rm R_i$ -- around unity.  Comparing with the results of the typical case (Figure \ref{fig:1e4compare}) shows that the abundances can be strongly affected by the temperature history of the disc.

\section{Discussion and comparison with observations}

To compare our results to the element compositions of the CM, CO, and CV chondrites we adopt the decoupling timescale of $1.5\times10^4$ yr and disc evolution timescale $t\rm{_e}=2.625\times10^4$ yr (identical to \citet{Cassen:1996}).  MC model reproduces the general trends for volatile elements of the chondrites as do our two new models.  However, the relative abundances are different at different radii.  Our full condensation model most closely matches the observed abundances (CM, CO, and CV) at 3 AU, i.e., consistent with the values of the relative abundances of the chondrites.

The more refractory elements of CM, CO, and CV chondrites with $\rm T_{50}$ higher than that for Ti (1583 K) have lower relative abundances (Figure \ref{fig:1e4compare}).  With our nominal decoupling and disc evolution timescales, the quick depletion of the refractory elements affects the transfer of elements to different radii.  This effect can explain the low relative abundances of the most refractory elements.  For example, the three most refractory elements Re, Os, and W have lower abundances compared to Ti in chondrites.

To the extent that the composition of the condensed materials reflects the composition of the terrestrial planets, our results also match the general trends for those planets. The average values of $\rm R_i$ for refractory elements for Mercury, Venus, Earth, and Mars are 1.80, 1.17, 1.32, and 1.02, respectively.  For volatile elements, the values are 0.0877, 0.274, 0.105, and 0.381 (See Figure \ref{fig:all-ele}).  Generally, the values for the refractories are high at small orbital distances (except Venus) and low for volatile elements -- consistent with the results shown in Figure \ref{fig:1e4compare}.

Figure \ref{fig:cassen-misha-final1e53} (b) shows that $\rm R_i$ for refractory elements can decrease toward small radii while increasing slightly for volatile elements.  A short decoupling timescale may explain the slightly lower refractory abundances of Venus compared to the Earth.  The surface density and the temperature in the inner region of the model are higher than what we would expect from a more realistic disc model \citep{Li:2017}.  For example, at one disc evolution time, the midplane temperature at 2 AU is around 1000 K, while at the same time, but using the disc model from \citet{Li:2017}, 1000 K occurs near 1 AU.  Therefore, if we obtain the same relative abundance for the terrestrial planets in our disc, the semimajor axes from our model will be larger than the actual locations of the planets in a real system.

Observations suggest that chondrule formation began after CAI formation and persisted for a few million years \citep{Kita:2000,Amelin:2002,Connelly:2012}.  
Our models yield chemical compositions of the condensed materials, reproducing the elemental depletion patterns in some chondrites, but they do not address the formation of either CAIs or chondrules.  Accounting for the differences between the formation of CAIs and chondrites would require simulating the evolution of the disc for several million years to track how the environment in the disc changes with time -- roughly two orders-of-magnitude beyond our simulation times.  In addition, since the pattern of the abundance of elements in CM, CO, and CV chondrites are measured from cm-scale samples \citep{Wood:2005,Ciesla:2008}, one should include some details of the planetesimal formation process 
in order to make more detailed comparisons with solar system planetesimals.

\section{Conclusions}

We present two condensation models that predict the chemical compositions of the chondrites (CM, CO, and CV) and terrestrial planets.
The first model follows \citet{Cassen:1996} by using $\rm T_{50}$ as a fiducial value below which the elements will begin to decouple.  In the second model, we calculate the chemical equilibrium of 33 elements by minimizing the Gibbs free energy of the system at each radius and specific time steps in the evolving disc.

Compared with previous works, our newest model improves upon them by 1) using a chemical equilibrium calculation to model the condensation of the elements rather than fixed $\rm T_{50}$, 2) significantly increasing the number of elements used in the calculation, and 3) embedding the elements in a dynamically and thermally evolving protoplanetary disc rather than choosing one or several snapshots in the evolving discs.

From our results we draw the following conclusions:

\begin{enumerate}[leftmargin=*]

\item For each element, the condensation front expands inward and outward from its initial condensation region.  The inner edge of this region advances because the temperature decreases with time.  The outer edge of the region retreats because of the outward movement of the gas in the disc, which increases the disc surface density.

\item The relative abundances of the condensed elements in the disc vary with time and distance, decreasing for refractory elements and increasing for volatile elements within $\sim$ 3 AU.  The resulting condensed materials are enriched in refractory and depleted in volatile elements within $\sim$ 3 AU.  The chemical compositions of chondrites (CM, CO and CV) are most closely matched by the condensate formed at 3 AU in these models.

\item Approximating the condensation process with a fixed $\rm T_{50}$ is similar to the full chemical equilibrium model.  However, there are systematic differences from refractory to volatile elements.  The relative elemental abundances, $\rm R_i$, calculated from the chemical equilibrium are higher for refractory and lower for volatile elements within 3 AU.  That is, the 50\% condensation model underestimates refractories and overestimates volatiles. Nevertheless, under certain circumstances, the $\rm T_{50}$ model may be sufficient. The systematic differences can be as large as 10\% and may become more important as disc models are further refined.

\item The abundances do not change significantly when the decoupling timescale is equal to or longer than the disc evolution timescale.  Rather, the surface density of the material decreases because advected dust drifts out of the region of interest.

\item If the decoupling time is short, there will be peaks in the elemental distribution as a function of the condensation temperature.  The relative abundance of refractory elements are diminished at small semimajor axis because refractory elements from more distant regions decouple while the volatile elements will flow toward the star, diluting the refractory elements.  This may explain the estimated lower abundance of refractory elements in Venus compared to Earth.

\item When the initial temperature in the disc is high, the relative abundances of refractory elements are above unity, while these are low for volatile elements.  
When the initial temperature is low, the relative abundances of refractory elements are around unity since they condense at the same time as Si.  Thus, we expect the relative elemental abundances, $\rm R_i$, to be affected by the temperature history of the disc.  

\end{enumerate}

Using more elements in the calculations allows a more detailed comparison of with chondrite composition.  In addition, by including the effects of the transport of matter in the disc, our results can provide more information on the history of dust within the disc and may provide better estimates for the dust in planet forming discs generally. 
Our model offers one explanation for both the decreasing of $\rm R_i$ with $\rm T_{50}$ and the cases that do not follow these trends.
We notice that the thermal history can have a significant effect on the results. By including this aspect of the evolution of the disc we can improve our understanding of chemical compositions of the meteorites and planets.

Some limitations to our current disc model can be improved upon with future work.  The surface density and the temperature in the inner region of our disc are high compared to a more realistic model \citep[e.g.][]{Li:2017}.  An improved model will result in a smaller semimajor axes for the formation location of the planetesimals and planets and may affect the relative elemental abundances.

\section*{Acknowledgements}

The authors thank the referee for constructive comments that improved this manuscript.  JHS, ML, and SH are supported by the NASA grant NNX16AK08G and NSF grant AST-1910955.  ZZ acknowledges support from the National Science Foundation under CAREER Grant Number AST-1753168 and Sloan Research Fellowship.




\bibliographystyle{mnras}
\bibliography{dust}

\begin{thebibliography}{}
\makeatletter
\relax
\def\mn@urlcharsother{\let\do\@makeother \do\$\do\&\do\#\do\^\do\_\do\%\do\~}
\def\mn@doi{\begingroup\mn@urlcharsother \@ifnextchar [ {\mn@doi@}
  {\mn@doi@[]}}
\def\mn@doi@[#1]#2{\def\@tempa{#1}\ifx\@tempa\@empty \href
  {http://dx.doi.org/#2} {doi:#2}\else \href {http://dx.doi.org/#2} {#1}\fi
  \endgroup}
\def\mn@eprint#1#2{\mn@eprint@#1:#2::\@nil}
\def\mn@eprint@arXiv#1{\href {http://arxiv.org/abs/#1} {{\tt arXiv:#1}}}
\def\mn@eprint@dblp#1{\href {http://dblp.uni-trier.de/rec/bibtex/#1.xml}
  {dblp:#1}}
\def\mn@eprint@#1:#2:#3:#4\@nil{\def\@tempa {#1}\def\@tempb {#2}\def\@tempc
  {#3}\ifx \@tempc \@empty \let \@tempc \@tempb \let \@tempb \@tempa \fi \ifx
  \@tempb \@empty \def\@tempb {arXiv}\fi \@ifundefined
  {mn@eprint@\@tempb}{\@tempb:\@tempc}{\expandafter \expandafter \csname
  mn@eprint@\@tempb\endcsname \expandafter{\@tempc}}}

\bibitem[\protect\citeauthoryear{{Amelin}, {Krot}, {Hutcheon}  \&
  {Ulyanov}}{{Amelin} et~al.}{2002}]{Amelin:2002}
{Amelin} Y.,  {Krot} A.~N.,  {Hutcheon} I.~D.,   {Ulyanov} A.~A.,  2002,
  \mn@doi [Science] {10.1126/science.1073950}, \href
  {https://ui.adsabs.harvard.edu/abs/2002Sci...297.1678A} {297, 1678}

\bibitem[\protect\citeauthoryear{{Asplund}, {Grevesse}  \& {Sauval}}{{Asplund}
  et~al.}{2005}]{Asplund:2005}
{Asplund} M.,  {Grevesse} N.,   {Sauval} A.~J.,  2005, in {Barnes} III T.~G.,
  {Bash} F.~N.,  eds,  Astronomical Society of the Pacific Conference Series
  Vol. 336, Cosmic Abundances as Records of Stellar Evolution and
  Nucleosynthesis. p.~25

\bibitem[\protect\citeauthoryear{{Asplund}, {Grevesse}, {Sauval}  \&
  {Scott}}{{Asplund} et~al.}{2009}]{Asplund:2009}
{Asplund} M.,  {Grevesse} N.,  {Sauval} A.~J.,   {Scott} P.,  2009, \mn@doi
  [\araa] {10.1146/annurev.astro.46.060407.145222}, \href
  {https://ui.adsabs.harvard.edu/abs/2009ARA&A..47..481A} {47, 481}

\bibitem[\protect\citeauthoryear{{Bond}, {Lauretta}  \& {O'Brien}}{{Bond}
  et~al.}{2010a}]{Bond:2010a}
{Bond} J.~C.,  {Lauretta} D.~S.,   {O'Brien} D.~P.,  2010a, \mn@doi [Icarus]
  {10.1016/j.icarus.2009.07.037}, \href
  {http://adsabs.harvard.edu/abs/2010Icar..205..321B} {205, 321}

\bibitem[\protect\citeauthoryear{{Bond}, {O'Brien}  \& {Lauretta}}{{Bond}
  et~al.}{2010b}]{Bond:2010b}
{Bond} J.~C.,  {O'Brien} D.~P.,   {Lauretta} D.~S.,  2010b, \mn@doi [The
  Astrophysical Journal] {10.1088/0004-637X/715/2/1050}, \href
  {http://adsabs.harvard.edu/abs/2010ApJ...715.1050B} {715, 1050}

\bibitem[\protect\citeauthoryear{{Cassen}}{{Cassen}}{1994}]{Cassen:1994}
{Cassen} P.,  1994, \mn@doi [\icarus] {10.1006/icar.1994.1195}, \href
  {http://ads.bao.ac.cn/abs/1994Icar..112..405C} {112, 405}

\bibitem[\protect\citeauthoryear{{Cassen}}{{Cassen}}{1996}]{Cassen:1996}
{Cassen} P.,  1996, \mn@doi [Meteoritics and Planetary Science]
  {10.1111/j.1945-5100.1996.tb02114.x}, \href
  {http://adsabs.harvard.edu/abs/1996M%26PS...31..793C} {31, 793}

\bibitem[\protect\citeauthoryear{{Cassen}}{{Cassen}}{2001}]{Cassen:2001}
{Cassen} P.,  2001, \mn@doi [Meteoritics and Planetary Science]
  {10.1111/j.1945-5100.2001.tb01908.x}, \href
  {https://ui.adsabs.harvard.edu/abs/2001M%26PS...36..671C} {36, 671}

\bibitem[\protect\citeauthoryear{{Ciesla}}{{Ciesla}}{2008}]{Ciesla:2008}
{Ciesla} F.~J.,  2008, \mn@doi [Meteoritics and Planetary Science]
  {10.1111/j.1945-5100.2008.tb00675.x}, \href
  {https://ui.adsabs.harvard.edu/abs/2008M%26PS...43..639C} {43, 639}

\bibitem[\protect\citeauthoryear{{Connelly}, {Bizzarro}, {Krot}, {Nordlund},
  {Wielandt}  \& {Ivanova}}{{Connelly} et~al.}{2012}]{Connelly:2012}
{Connelly} J.~N.,  {Bizzarro} M.,  {Krot} A.~N.,  {Nordlund} {\AA}.,
  {Wielandt} D.,   {Ivanova} M.~A.,  2012, \mn@doi [Science]
  {10.1126/science.1226919}, \href
  {https://ui.adsabs.harvard.edu/abs/2012Sci...338..651C} {338, 651}

\bibitem[\protect\citeauthoryear{{Elser}, {Meyer}  \& {Moore}}{{Elser}
  et~al.}{2012}]{Elser:2012}
{Elser} S.,  {Meyer} M.~R.,   {Moore} B.,  2012, \mn@doi [\icarus]
  {10.1016/j.icarus.2012.09.016}, \href
  {http://adsabs.harvard.edu/abs/2012Icar..221..859E} {221, 859}

\bibitem[\protect\citeauthoryear{{Grossman}}{{Grossman}}{1972}]{Grossman:1972}
{Grossman} L.,  1972, \mn@doi [\gca] {10.1016/0016-7037(72)90078-6}, \href
  {http://adsabs.harvard.edu/abs/1972GeCoA..36..597G} {36, 597}

\bibitem[\protect\citeauthoryear{{Jacquet}, {Gounelle}  \& {Fromang}}{{Jacquet}
  et~al.}{2012}]{Jacquet:2012}
{Jacquet} E.,  {Gounelle} M.,   {Fromang} S.,  2012, \mn@doi [\icarus]
  {10.1016/j.icarus.2012.04.022}, \href
  {https://ui.adsabs.harvard.edu/abs/2012Icar..220..162J} {220, 162}

\bibitem[\protect\citeauthoryear{{Kita}, {Nagahara}, {Togashi}  \&
  {Morishita}}{{Kita} et~al.}{2000}]{Kita:2000}
{Kita} N.~T.,  {Nagahara} H.,  {Togashi} S.,   {Morishita} Y.,  2000, \mn@doi
  [\gca] {10.1016/S0016-7037(00)00488-9}, \href
  {https://ui.adsabs.harvard.edu/abs/2000GeCoA..64.3913K} {64, 3913}

\bibitem[\protect\citeauthoryear{{Larimer}}{{Larimer}}{1967}]{Larimer:1967}
{Larimer} J.~W.,  1967, \mn@doi [\gca] {10.1016/S0016-7037(67)80013-9}, \href
  {https://ui.adsabs.harvard.edu/abs/1967GeCoA..31.1215L} {31, 1215}

\bibitem[\protect\citeauthoryear{{Lewis}}{{Lewis}}{1974}]{Lewis:1974}
{Lewis} J.~S.,  1974, \mn@doi [Science] {10.1126/science.186.4162.440}, \href
  {http://http://adsabs.harvard.edu/abs/1974Sci...186..440L} {186, 440}

\bibitem[\protect\citeauthoryear{{Li} \& {Sui}}{{Li} \& {Sui}}{2017}]{Li:2017}
{Li} M.,  {Sui} N.,  2017, \mn@doi [\mnras] {10.1093/mnras/stw3200}, \href
  {http://adsabs.harvard.edu/abs/2017MNRAS.466.1205L} {466, 1205}

\bibitem[\protect\citeauthoryear{{Lodders}}{{Lodders}}{2003}]{Lodders:2003}
{Lodders} K.,  2003, \mn@doi [\apj] {10.1086/375492}, \href
  {http://adsabs.harvard.edu/abs/2003ApJ...591.1220L} {591, 1220}

\bibitem[\protect\citeauthoryear{{Lodders} \& {Fegley}}{{Lodders} \&
  {Fegley}}{1997}]{Lodders:1997}
{Lodders} K.,  {Fegley} B.,  1997, \mn@doi [\icarus] {10.1006/icar.1996.5653},
  \href {http://adsabs.harvard.edu/abs/1997Icar..126..373L} {126, 373}

\bibitem[\protect\citeauthoryear{{Lord}}{{Lord}}{1965}]{Lord:1965}
{Lord} III H.~C.,  1965, \mn@doi [\icarus] {10.1016/0019-1035(65)90005-9},
  \href {https://ui.adsabs.harvard.edu/abs/1965Icar....4..279L} {4, 279}

\bibitem[\protect\citeauthoryear{{McDonough} \& {Sun}}{{McDonough} \&
  {Sun}}{1995}]{McDonough:1995}
{McDonough} W.~F.,  {Sun} S.-s.,  1995, \mn@doi [Chemical Geology]
  {10.1016/0009-2541(94)00140-4}, \href
  {http://adsabs.harvard.edu/abs/1995ChGeo.120..223M} {120, 223}

\bibitem[\protect\citeauthoryear{{Morgan} \& {Anders}}{{Morgan} \&
  {Anders}}{1980}]{Morgan:1980}
{Morgan} J.~W.,  {Anders} E.,  1980, \mn@doi [Proceedings of the National
  Academy of Science] {10.1073/pnas.77.12.6973}, \href
  {http://adsabs.harvard.edu/abs/1980PNAS...77.6973M} {77, 6973}

\bibitem[\protect\citeauthoryear{{Moriarty}, {Madhusudhan}  \&
  {Fischer}}{{Moriarty} et~al.}{2014}]{Moriarty:2014}
{Moriarty} J.,  {Madhusudhan} N.,   {Fischer} D.,  2014, \mn@doi [\apj]
  {10.1088/0004-637X/787/1/81}, \href
  {http://ads.bao.ac.cn/abs/2014ApJ...787...81M} {787, 81}

\bibitem[\protect\citeauthoryear{{O'Brien}, {Morbidelli}  \&
  {Levison}}{{O'Brien} et~al.}{2006}]{OBrien:2006}
{O'Brien} D.~P.,  {Morbidelli} A.,   {Levison} H.~F.,  2006, \mn@doi [\icarus]
  {10.1016/j.icarus.2006.04.005}, \href
  {http://ads.bao.ac.cn/abs/2006Icar..184...39O} {184, 39}

\bibitem[\protect\citeauthoryear{{Palme}, {Lodders}  \& {Jones}}{{Palme}
  et~al.}{2014}]{Palme:2014}
{Palme} H.,  {Lodders} K.,   {Jones} A.,  2014, {Solar System Abundances of the
  Elements}.
pp 15--36

\bibitem[\protect\citeauthoryear{Petaev}{Petaev}{2009}]{Petaev:2009}
Petaev M.~I.,  2009, \mn@doi [Calphad] {10.1016/j.calphad.2008.12.001}, 33, 317

\bibitem[\protect\citeauthoryear{{Pignatale}, {Liffman}, {Maddison}  \&
  {Brooks}}{{Pignatale} et~al.}{2016}]{Pignatale:2016}
{Pignatale} F.~C.,  {Liffman} K.,  {Maddison} S.~T.,   {Brooks} G.,  2016,
  \mn@doi [\mnras] {10.1093/mnras/stv3003}, \href
  {http://adsabs.harvard.edu/abs/2016MNRAS.457.1359P} {457, 1359}

\bibitem[\protect\citeauthoryear{{Urey}}{{Urey}}{1955}]{Urey:1955}
{Urey} H.~C.,  1955, \mn@doi [Proceedings of the National Academy of Science]
  {10.1073/pnas.41.3.127}, \href
  {https://ui.adsabs.harvard.edu/abs/1955PNAS...41..127U} {41, 127}

\bibitem[\protect\citeauthoryear{{Wasson} \& {Kallemeyn}}{{Wasson} \&
  {Kallemeyn}}{1988}]{Wasson:1988}
{Wasson} J.~T.,  {Kallemeyn} G.~W.,  1988, \mn@doi [Philosophical Transactions
  of the Royal Society of London Series A] {10.1098/rsta.1988.0066}, \href
  {http://adsabs.harvard.edu/abs/1988RSPTA.325..535W} {325, 535}

\bibitem[\protect\citeauthoryear{{Weidenschilling} \&
  {Cuzzi}}{{Weidenschilling} \& {Cuzzi}}{1993}]{Weidenschilling:1993}
{Weidenschilling} S.~J.,  {Cuzzi} J.~N.,  1993, \href
  {https://ui.adsabs.harvard.edu/abs/1993prpl.conf.1031W} {p.~1031}

\bibitem[\protect\citeauthoryear{{Woitke}, {Helling}, {Hunter}, {Millard},
  {Turner}, {Worters}, {Blecic}  \& {Stock}}{{Woitke}
  et~al.}{2018}]{Woitke:2018}
{Woitke} P.,  {Helling} C.,  {Hunter} G.~H.,  {Millard} J.~D.,  {Turner} G.~E.,
   {Worters} M.,  {Blecic} J.,   {Stock} J.~W.,  2018, \mn@doi [\aap]
  {10.1051/0004-6361/201732193}, \href
  {https://ui.adsabs.harvard.edu/abs/2018A&A...614A...1W} {614, A1}

\bibitem[\protect\citeauthoryear{{Wood}}{{Wood}}{2005}]{Wood:2005}
{Wood} J.~A.,  2005, in {Krot} A.~N.,  {Scott} E.~R.~D.,   {Reipurth} B.,  eds,
   Astronomical Society of the Pacific Conference Series Vol. 341, Chondrites
  and the Protoplanetary Disk. p.~953

\makeatother
\end{thebibliography}




\appendix




\bsp	
\label{lastpage}
\end{document}